\def\nbh{{\rm n}}
\def\acap{\\ \nonumber \\}

\def\nnb{\nonumber }

\def\sif{\sin f}
\def\cif{\cos f}

\def\kx{{{\hat{S}}_x}}
\def\ky{{{\hat{S}}_y}}
\def\kz{{{\hat{S}}_z}}

\def\nk{n_{\rm b}}

\def\rfr#1{eq. (\ref{#1})}

\def\derp#1#2{\rp{\partial{#1}}{\partial{#2}}}
\def\dert#1#2{\frac{{{d}}{#1}}{{{d}}{#2}}}

\def\virg#1{``#1"}

\def\eqi{\begin{equation}}
\def\eqf{\end{equation}}
\def\eqia{\begin{eqnarray}}
\def\eqfa{\end{eqnarray}}
\def\Om{\mathit{\Omega}}
\def\rp#1#2{{#1\over#2}}
\def\lb#1{\label{#1}}
\def\kap{\bds{\hat{S}}}

\def\bds#1{\boldsymbol{#1}}

\def\co{\cos\omega}
\def\so{\sin\omega}
\def\coo{\cos 2\omega}
\def\soo{\sin 2\omega}
\def\cO{\cos\Om}
\def\sO{\sin\Om}
\def\cOO{\cos 2\Om}
\def\sOO{\sin 2\Om}

\def\cI{\cos I}
\def\sI{\sin I}
\def\cII{\cos 2I}
\def\sII{\sin 2I}

\def\ton#1{\left(#1\right)}
\def\qua#1{\left[#1\right]}
\def\grf#1{\left\{#1\right\}}

\documentclass[11pt]{article}

\usepackage{lscape,rotating,amsmath,amssymb,amsthm,amscd,latexsym,wasysym}
\usepackage[toc,title,titletoc]{appendix}
\usepackage{cite}
\usepackage{graphicx,epsfig,textgreek,w-greek}
\usepackage[unicode,breaklinks,
colorlinks=true,
linkcolor=red,
urlcolor=blue]{hyperref}
\usepackage{breakurl}

\RequirePackage{color}

\allowdisplaybreaks[1]

\RequirePackage{color}

\begin{document}

\title{Post-Newtonian direct and mixed orbital effects due to the oblateness of the central body}

\author{L. Iorio\\ CNR-Istituto di metodologie inorganiche e dei plasmi (I.M.I.P)\\ Via Amendola 122/D, 70126, Bari (BA), Italy\\ email: lorenzo.iorio@libero.it}

\maketitle

\begin{abstract}
The orbital dynamics of a test particle moving in the non-spherically symmetric
field of a rotating oblate primary is impacted also by certain
indirect, mixed effects arising from the interplay of the different Newtonian and post-Newtonian accelerations
which induce known direct perturbations.
We systematically calculate the indirect gravitoelectromagnetic
shifts per orbit of the Keplerian orbital elements of the test
particle arising from the crossing among the first even zonal harmonic $J_2$ of the central
body and the post-Newtonian static and stationary components
of its gravitational field. We also work out the Newtonian shifts per
orbit of order $J_2^2$, and the direct post-Newtonian gravitoelectric  effects of order
$J_2 c^{-2}$ arising from the equations of motion. In the case of
both the indirect and direct gravitoelectric $J_2 c^{-2}$ shifts, our calculation
holds for an arbitrary orientation of the symmetry axis of the central
body. We yield numerical estimates of their relative magnitudes for
systems ranging from Earth artificial satellites to stars orbiting supermassive
black holes.
\textcolor{black}{As far as their measurability is concerned, highly elliptical orbital configurations are desirable.}
\end{abstract}

\section{Introduction}
Accurate tests of post-Newtonian gravity with either natural and artificial probes \cite{Ginz59, 2010stsp.conf...25O, 2013CEJPh..11..531R, 2013NuPhS.243..172W} in a variety of astronomical and astrophysical scenarios as well as the long-term dynamics of hierarchical systems \cite{2014PhRvD..89d4043W} require an ever increasing accuracy in modeling their orbital dynamics. In this respect, first-order perturbative calculations, yielding some of the most renown direct orbital consequences of the equations of motion \cite{Sof89} like the Einstein perihelion precession \cite{Ein15} and the Lense-Thirring effect \cite{LT18},  should be complemented by second-order calculations. Indeed, they are able to capture also certain subtle consequences of the simultaneous presence of several terms in the equations of motion. They are indirect, mixed effects arising from the interplay among such terms which, in some cases, may have magnitudes comparable to some of the direct effects to the point that it has been recently stressed that they might be the subject of promising experimental investigations in a near future \cite{2013CQGra..30s5011I}.
While it is assumed that the orbital elements stay constant over one orbital revolution in calculating perturbatively the direct effects to the first order in some disturbing acceleration, accounting also for their instantaneous variations \cite{2014PhRvD..89d4043W} yields either second-order and mixed effects if the acceleration is, actually, made of the sum of more than one term.

More specifically, let us consider a test particle moving in the non-spherically symmetric field of a rotating, oblate primary of mass $M$, mean equatorial radius $R$, quadrupole moment $J_2$ and angular momentum $\bds S$: apart from the Newtonian monopole, the acceleration experienced by the particle (Section \ref{pertu}) consists of the sum of a Newtonian term ${\bds A}^{(J_2)}$ accounting for the primary's oblateness\footnote{Here and throughout the paper, the other even zonal coefficients $J_{\ell},\ell=4,6,\ldots$ of higher degree of the Newtonian multipolar expansion of the gravitational potential of the central body will be neglected.} $J_2$ and, to order $\mathcal{O}\ton{c^{-2}}$, of some static and stationary post-Newtonian terms ${\bds A}^{(\rm GE)},~{\bds A}^{(\rm GM)},~{\bds A}^{(J_2~{\rm GE})}$ yielding, to the first-order in each of them, direct effects like the gravitoelectric Schwarzschild-type  precession of the line of the apsides \cite{Ein15}, the gravitomagnetic Lense-Thirring orbital precessions \cite{LT18}, and some further orbital precessions proportional to $J_2 c^{-2}$ \cite{1988CeMec..42...81S, 1991ercm.book.....B}. Actually, a perturbative calculation accounting also for the instantaneous shifts of the orbital elements during an orbital revolution is able to deliver, among other things, also mixed effects among such accelerations. It should be recalled that, in the perturbed 2-body Newtonian scenario, the 2nd order short term effects could be larger than the secular ones \cite{2013osos.book.....X}. As a result, mixed orbital variations of order $\mathcal{O}\ton{J_2 c^{-2}}$, which are to be added to those directly arising from the equations of motion through ${\bds A}^{(J_2~{\rm GE})}$ \cite{1988CeMec..42...81S, 1991ercm.book.....B}, occur. As recognized long ago, \cite{Sof89, 1990CeMDA..47..205H}, they are of the same order of magnitude of the direct effects due to ${\bds A}^{(J_2~{\rm GE})}$. The gravitoelectric mixed effects were calculated in \cite{2014PhRvD..89d4043W}, although they were not explicitly displayed. A calculation of them, based on Lie series and the Delaunay variables, can be found in \cite{1990CeMDA..47..205H, arabo}. Moreover, there are also other mixed effects proportional to $J_2 S c^{-2}$ due to the interplay between the Newtonian quadrupolar field and the post-Newtonian gravitomagnetic one; to the best of our knowledge, they were never calculated in the literature. Direct orbital effects of order $\mathcal{O}\ton{J_2 S c^{-2}}$, calculated on the basis of a suitable multipolar expansion of the gravitomagnetic field of an axially symmetric source \cite{1978PhRvD..18.1037T}, can be found in \cite{1977PhRvD..16..946T}. For the direct post-Newtonian effects pertaining the precession of a gyroscope orbiting a rotating oblate body, see \cite{1977PhRvD..16..946T, 1999GReGr..31.1837A, 2008arXiv0803.4133Z, 2014CQGra..31x5012P}
As a by-product of such a calculation, also classical
orbital shifts of order $\mathcal{O}\ton{J_2^2}$
can be obtained.

In this paper, we will analytically work out all the aforementioned effects (Section \ref{misti}-Section \ref{J2J2}) in a systematic and consistent way, outlined in Section \ref{gene}, which, in principle, can be applied also to other disturbing accelerations, irrespectively of their physical origin.
As far as both the mixed and the direct effects proportional to $J_2c^{-2}$ are concerned, we will calculate them for an arbitrary orientation of the spin axis of the central body.
Our results, which are valid for generic orbital geometries of the test particle, represent the limits to which full two-body calculations must reduce in the point particle limit.  They can be used for sensitivity analyses involving different scenarios.
Then, in Section \ref{expe} we will numerically evaluate the relative strengths of such orbital shifts in various systems ranging from planetary artificial satellites \cite{2007AcAau..61..932M, 2011AcAau..69..127P} to the stellar system orbiting the supermassive black hole (BH) located in Sgr A$^{\ast}$ at the center of the Galaxy \cite{2008ApJ...689.1044G}. Section \ref{concludi} summarizes our findings.
\subsection*{Notations}
Here,  basic notations and definitions used in the text are presented \cite{1991ercm.book.....B, 2000Monte}.
\begin{description}
\item[] $G:$ Newtonian constant of gravitation
\item[] $c:$ speed of light in vacuum
\item[] $M:$ mass of the primary
\item[] $\mu=GM:$ gravitational parameter of the primary
\item[] ${\mathcal{R}}_s = 2\mu c^{-2}: $ Schwarzschild radius of the primary
\item[] $R:$ mean equatorial radius of the primary
\item[] $J_2:$ dimensionless quadrupole mass moment of the primary
\item[] $S:$ angular momentum of the primary
\item[] ${\kap}:$ unit vector of the spin axis of the primary
\item[] ${\bds r}:$ radius vector of the test particle
\item[] $\bds{\hat{r}} = \bds{r}/r:$ unit vector of the radius vector of the test particle
\item[] ${\bds v}:$ velocity vector of the test particle
\item[] $\bds k = \bds r\bds\times \bds v:$ orbital angular momentum per unit mass of the test particle
\item[] $\bds{\hat{k}}={\bds k}/k:$ unit vector of the orbital angular momentum per unit mass of the test particle
\item[] $a:$  semimajor axis
\item[] $\nk = \sqrt{\mu a^{-3}}:$   Keplerian mean motion
\item[] $P_{\rm b} = 2\pi \nk^{-1}:$ Keplerian orbital period
\item[] $e:$  eccentricity
\item[] $p=a(1-e^2):$  semilatus rectum
\item[] $I:$  inclination of the orbital plane
\item[] $\Om:$  longitude of the ascending node
\item[] $\omega:$  argument of pericenter
\item[] $f:$  true anomaly
\item[] $u=\omega + f:$  argument of latitude
\item[] $\bds{\hat{l}}:$ unit vector directed along the line of the nodes toward the ascending node
\item[] $\bds{\hat{m}}:$ unit vector directed transversely to the line of the nodes in the orbital plane
\item[] $\bds{\hat{P}}:$ unit vector directed along the line of the apsides toward the pericenter
\item[] $\bds{\hat{Q}}:$ unit vector directed transversely to the line of the apsides in the orbital plane
\item[] $\bds A:$ disturbing acceleration
\item[] $A_R=\bds A\bds\cdot{\bds{\hat{r}}}:$ radial component of $\bds A$
\item[] $A_T=\bds A\bds\cdot\ton{{\bds{\hat{k}}}\bds\times{\bds{\hat{r}}}}:$ transverse component of $\bds A$
\item[] $A_N=\bds A\bds\cdot{\bds{\hat{k}}}:$ normal component of $\bds A$
\end{description}
\section{General scheme for calculating the second-order and mixed orbital shifts}\lb{gene}
A suitable form of the Gauss equations \cite{2011MNRAS.410..654X, 2013MNRAS.432..584X, 2013osos.book.....X} for the variation of the Keplerian orbital elements in presence of a perturbing acceleration $\bds A$ is \cite{1959ForPh...7S..55T, 1961mcm..book.....B, 1970CeMec...2..369R, 1979AN....300..313M}
\begin{align}
\dert p f \lb{dpdf}& = \rp{2r^3\upgamma A_T}{\mu}, \\ \nonumber \\
\dert e f & = \rp{r^2\upgamma}{\mu}\qua{\sin f A_R + \ton{1+\rp{r}{p}}\cos f A_T  +e\ton{\rp{r}{p}}A_T},\\ \nonumber \\
\dert I f &= \rp{r^3\upgamma \cos u A_N}{\mu p }, \\ \nonumber \\
\dert \Om f &= \rp{r^3\upgamma \sin u A_N}{\mu p \sI}, \\ \nonumber \\
\dert\omega f \lb{dodf}& = \rp{r^2\upgamma}{\mu e}\qua{-\cos f A_R +\ton{1+\rp{r}{p} }\sin f A_T}-\cI\dert\Om f,\\ \nonumber \\
\dert t f &= \rp{r^2\upgamma}{\sqrt{\mu p}},
\end{align}
with \cite{1958SvA.....2..147E, 1959ForPh...7S..55T, 1961mcm..book.....B, 1970CeMec...2..369R, 1979AN....300..313M, 1991ercm.book.....B}
\eqi \upgamma  = \rp{1}{1-  \rp{r^2}{\sqrt{\mu p}}\ton{\dert\omega t + \cI\dert\Om t}} \lb{gamma}. \eqf
The $\upgamma$ factor arises because the true anomaly $f$ is recknoned from the pericenter position which, in general, is shifted by $\bds A$ through the changes of the longitude of the ascending node $\Om$ and the argument of pericenter $\omega$ occurring during an orbital revolution \cite{2014PhRvD..89d4043W}.
To the first order in the perturbation, $\upgamma$ can be expressed as
\begin{align}\upgamma \nonumber &\simeq 1 + \rp{r^2}{\sqrt{\mu p}}\ton{\dert\omega t +\cI\dert\Om t} =\\ \nonumber \\
\lb{gammaexp} &= 1+ \rp{r^2}{\mu e}\qua{-\cos f A_R +\ton{1+\rp{r}{p} }\sin f A_T}.
\end{align}

To the second order in $\bds A$, the angular rate of change of a generic Keplerian orbital element $\varphi_i,\ i=p,e,I,\Om,\omega$ can be expanded as\footnote{An analogous approach is followed in the literature \cite{1977AN....298..107M, 1979AN....300..313M} for $dt/df,\ dt/du$ to calculate the anomalistic and the draconitic perturbed periods to the first order in ${\bds A}$.}
\begin{align}\dert{\varphi_i} f \nonumber  \lb{dphidf}&= \grf{\dert{\varphi_i} f}_0 + \sum_{j=p,e,I,\Om,\omega}\grf{\derp{\ton{d\varphi_i/df}}{\varphi_j}}_0\Delta\varphi_j^{(1)}\ton{f_0,f}+\\ \nonumber \\
& + \grf{\dert{\varphi_i}f\rp{r^2}{\mu e}\qua{-\cos f A_R +\ton{1+\rp{r}{p} }\sin f A_T}}_0 + \cdots.
\end{align}
 In \rfr{dphidf}, the curly brackets $\grf{\cdots}_0$ denote that the quantity inside has to be evaluated onto the unperturbed Keplerian ellipse.
 The second term in \rfr{dphidf} accounts for the fact that, actually, all the orbital parameters slowly change during an orbital revolution because of
 $\bds A$ \cite{2014PhRvD..89d4043W}; in a standard first-order calculation, such variations are usually neglected by assuming that the Keplerian elements can be considered as fixed to some fiducial values at an epoch $t_0$.
 The instantaneous shifts in \rfr{dphidf} are  calculated as
\eqi \Delta\varphi_j^{(1)}\ton{f_0,f}=\int_{f_0}^f \dert{\varphi_j}{f^{'}}df^{'} \lb{Dphi}\eqf
by using \rfr{dpdf}-\rfr{dodf} with $\upgamma=1$; the superscript $(1)$ in \rfr{Dphi} denotes that they are accurate to the first order in the perturbing acceleration. By integrating \rfr{dphidf} from $f_0$ to $f_0+2\pi$ allows to obtain the shift per orbit\footnote{Actually, it could be defined as an anomalistic shift. Indeed, the variable of integration is the true anomaly $f$, so that it refers to two consecutive passages at the pericenter, which, in general, does not stay constant in presence of a perturbation.}  $\Delta\varphi_i^{(2)}$ accurate to the second order in  $\bds A$.

If the latter one can be expressed as the sum of two perturbations ${\bds A}^{\rm A}$ and ${\bds A}^{\rm B}$, the second and the third terms of \rfr{dphidf} yield both the quadratic changes $\Delta\varphi_i^{(2)}$ for each of the disturbing accelerations and the mixed shifts $\Delta\varphi_i^{(\rm AB)}$ due to both of them.
\section{The Newtonian and relativistic post-Keplerian disturbing accelerations and their first-order orbital shifts}\lb{pertu}
In perturbative calculations, the disturbing acceleration $\bds A$ is usually projected onto three mutually orthogonal directions; the resulting components are then evaluated onto the unperturbed Keplerian ellipse. Here, we outline the general features of the procedure which can be applied to any perturbation, independently of its physical origin.

In this Section, we will treat the most important Newtonian and relativistic post-Keplerian accelerations by deriving also the instantaneous variations of the orbital elements induced by them. Such expressions, which will be used in Section \ref{misti}-Section \ref{J2J2} in calculating the mixed and second-order effects, are also important \textit{per se}  because the characteristic timescales of several astronomical and astrophysical scenarios of potential interest are quite longer than the observational time spans during which data are usually collected. Suffice it to say that, e.g., modern astrometric observations do not yet cover a full orbital revolution of Neptune \cite{2009ApJ...697.1226P}. About the stellar system orbiting the supermassive BH in Sgr A$^{\ast}$, observations spanning at least an orbital period have been collected so far only for two stars \cite{2008ApJ...689.1044G, 2009ApJ...692.1075G, 2012Sci...338...84M, 2014IAUS..303..242B}. Currently available data for, say, the Magellanic clouds necessarily refer to a tiny fraction of their orbital revolutions about the Galaxy \cite{2006ApJ...652.1213K, 2006ApJ...638..772K, 2011MNRAS.413.2015D}; general relativity has also been proposed to explain the Dark Matter-related issue of the galactic rotation curves \cite{2006MPLA...21.2133C, 2008MPLA...23.1745C, 2012Ap&SS.337..321C}. Thus, knowing accurately also such short-term effects is important because, over timescales quite shorter than the orbital periods of the systems considered, they may look as long-term, semi-secular signatures, somewhat aliasing the recovery of the genuine secular trends of interest \cite{2011MNRAS.415.3335X}.

The components of the unit vectors $\bds{\hat{l}},\bds{\hat{m}},\bds{\hat{k}}$ are \cite{1991ercm.book.....B}
\begin{align}
{\hat{l}}_x & =\cO, \\ \nonumber \\
{\hat{l}}_y & =\sO, \\ \nonumber \\
{\hat{l}}_z & =0, \\ \nonumber \\
{\hat{m}}_x & = -\cI\sO, \\ \nonumber \\
{\hat{m}}_y & = \cI\cO, \\ \nonumber \\
{\hat{m}}_z & = \sI, \\ \nonumber \\
{\hat{k}}_x & =\sI\sO, \\ \nonumber \\
{\hat{k}}_y & = -\sI\cO, \\ \nonumber \\
{\hat{k}}_z & = \cI,
\end{align}
so that it is \cite{1991ercm.book.....B}
\begin{align}
\bds{\hat{P}} & = \bds{\hat{l}}\co + \bds{\hat{m}}\so, \\ \nonumber \\
\bds{\hat{Q}} & = -\bds{\hat{l}}\so + \bds{\hat{m}}\co.
\end{align}
Thus, the position vector can be expressed as \cite{1991ercm.book.....B}
\eqi
\bds r  = r\ton{\bds{\hat{P}}\cos f + \bds{\hat{Q}}\sin f},\lb{vecr}
\eqf
with
\eqi r = \rp{p}{1+e\cos f}. \eqf
Moreover, the velocity vector is \cite{1991ercm.book.....B}
\eqi
\bds v  =  \sqrt{\rp{\mu}{p}}\qua{-\bds{\hat{P}}\sin f + \bds{\hat{Q}}\ton{\cos f + e}}.
\eqf

The radial, transverse and normal components of $\bds A$ can be finally calculated as \cite{1991ercm.book.....B}
\begin{align}
A_R & =\bds A\bds\cdot \bds{\hat{r}}, \\ \nonumber \\
A_T & = \bds A\bds\cdot \ton{\bds{\hat{k}}\bds\times\bds{\hat{r}} }, \\ \nonumber \\
A_N & = \bds A\bds\cdot \bds{\hat{k}}.
\end{align}
\subsection{The post-Newtonian gravitoelectric acceleration}\lb{ge}
The post-Newtonian gravitoelectric acceleration due to a static mass $M$ is \cite{Sof89}
\eqi{\bds A}^{(\rm GE)} = -\rp{\mu}{c^2 r^2}\ton{v^2-\rp{4\mu}{r}}\bds{\hat{r}} +\rp{4\mu}{c^2 r^2}\ton{\bds{\hat{r}}\bds\cdot\bds v}\bds v.\lb{Schwacc}\eqf
Its radial, transverse and normal components are \cite{Sof89}
\begin{align}
A_R^{(\rm GE)} \lb{ARGE} & =\rp{\mu^2\ton{1+e\cos f}^2\ton{3 + e^2 + 2e\cos f - 2e^2\cos 2f}}{c^2 p^3 }, \\ \nonumber \\
A_T^{(\rm GE)} \lb{ATGE}& = \rp{4\mu^2\ton{1+e\cos f}^3 e\sin f}{c^2 p^3 }, \\ \nonumber \\
A_N^{(\rm GE)} \lb{ANGE}&= 0.
\end{align}
The instantaneous shifts of the orbital elements, calculated as in \rfr{Dphi}, are
\begin{align}
\Delta p^{(\rm GE)}\ton{f,f_0}  \lb{pGE} & = \rp{8e\mu\ton{\cos f_0 -\cos f}}{c^2}, \\ \nonumber \\
\Delta e^{(\rm GE)}\ton{f,f_0}  & = \rp{\mu\ton{\cos f_0 -\cos f}\qua{3+7e^2+5e\ton{\cos f_0 +\cos f} }}{c^2 p}, \\ \nonumber \\
\Delta I^{(\rm GE)}\ton{f,f_0}  & = 0, \\ \nonumber \\
\Delta \Om^{(\rm GE)}\ton{f,f_0}  & = 0, \\ \nonumber \\
\Delta \omega^{(\rm GE)}\ton{f,f_0}  \nonumber \lb{oGE} & = \rp{\mu}{c^2 ep}\qua{3e\ton{f-f_0} +\ton{3-e^2 +5e\cos f_0}\sin f_0  -\right.\\ \nonumber \\
& -\left. \ton{3-e^2 +5e\cos f}\sin f  }.
\end{align}
Over one orbit, \rfr{pGE}-\rfr{oGE} yield the shifts
\begin{align}
\Delta p^{\ton{\rm GE}} & = 0, \acap
\Delta e^{\ton{\rm GE}} & = 0, \acap
\Delta I^{\ton{\rm GE}} & = 0, \acap
\Delta \Om^{\ton{\rm GE}} & = 0, \acap
\Delta \omega^{\ton{\rm GE}} \lb{omegaGE} & = \rp{6\pi\mu}{c^2 p}.
\end{align}
\subsection{The post-Newtonian gravitomagnetic  acceleration}\lb{gvm}
The Lense-Thirring acceleration  experienced by a test particle orbiting in the stationary gravitomagnetic field of a  rotating primary  is \cite{2010ITN....36....1P}
\eqi
{\bds A}^{(\rm GM)} \lb{ALT}  = \rp{2GS}{c^2 r^3}\qua{\rp{3(\kap\bds\cdot\bds r)\bds r\bds\times\bds v}{r^2}+ \bds v\bds\times\kap}.
\eqf

For a generic orientation of the spin axis $\kap$ of the central body, the methods reviewed in Section \ref{pertu} applied to \rfr{ALT} yield
\begin{align}
A_R^{\rm (GM)} \lb{ALTR} \nonumber & = \rp{2GS\sqrt{\mu}\ton{1+e\cif}^4}{c^2 \sqrt{p^7}}\qua{\kz\cI+\right. \\ \nonumber \\
&+\left. \sI\ton{\kx\sO-\ky\cO} }, \\ \nonumber \\
A_T^{\rm (GM)} \lb{ALTT} \nonumber & = -\rp{2eGS\sqrt{\mu}\ton{1+e\cif}^3\sif}{c^2 \sqrt{p^7}}\qua{\kz\cI+\right.\\ \nonumber \\
                         &+\left. \sI\ton{\kx\sO-\ky\cO} }, \\ \nonumber \\
A_N^{\rm (GM)} \nonumber \lb{ALTN} & = -\rp{GS\sqrt{\mu}\ton{1+e\cif}^3}{c^2 \sqrt{p^7}}\grf{-\kz\sI\qua{ e\so + \right.\right. \\ \nonumber \\
\nonumber &+\left.\left. 4\sin u  + 3e\sin\ton{2f +\omega} }  + \cI\qua{e\so + 4\sin u  + \right.\right. \\ \nonumber \\
\nonumber &+\left.\left. 3e\sin\ton{2f +\omega}}\ton{\kx\sO -\ky\cO} - \right.\\ \nonumber \\
          & - \left. \qua{e\co + 4\cos u  + 3e\cos\ton{2f +\omega}}\ton{\kx\cO + \ky\sO}}.
\end{align}
In the particular case $\kx=\ky=0,\kz=1$, \rfr{ALTR}-\rfr{ALTN} agree with Eq. (4.2.18a)- Eq. (4.2.18c) in \cite{Sof89}.

The instantaneous shifts due to \rfr{ALT}, calculated as in \rfr{Dphi}, are too cumbersome to be displayed for a generic orientation of $\kap$.
For $\kx=\ky=0,\kz=1$, they are
\begin{align}
\Delta p^{\ton{\rm GM}}\ton{f,f_0} \lb{plt} & = \rp{4GSe\cI\ton{\cos f - \cos f_0}}{c^2\sqrt{p\mu}}, \acap
\Delta e^{\ton{\rm GM}}\ton{f,f_0} \lb{elt} & = -\rp{2GS\ton{1-e^2}\cI\ton{\cos f - \cos f_0}}{c^2\sqrt{p^3\mu}}, \acap
\Delta I^{\ton{\rm GM}}\ton{f,f_0} \nnb \lb{Ilt} & = -\rp{GS\sI}{c^2\sqrt{p^3\mu}}\grf{
2 e \cos f  \cos ^2 u - \right.\acap
     & - \left. 2 e \cos  f_0  \cos ^2 u_0  + \cos 2u - \cos 2u_0 }, \acap
\Delta \Om^{\ton{\rm GM}}\ton{f,f_0} \nnb \lb{Olt} & = \rp{GS}{c^2\sqrt{p^3\mu}}\grf{
2f - 2 f_0 + 2 e \sin f - 2 e \sin  f_0 - \right.\acap
     & - \left.\ton{1 + e \cos f} \sin 2u + \ton{1 + e \cos  f_0}\sin 2u_0 }, \acap
\Delta \omega^{\ton{\rm GM}}\ton{f,f_0} \nnb \lb{omlt} & = -\rp{GS\cI}{c^2 e\sqrt{p^3\mu}}\grf{
\left(2 + 4 e^2\right) \sin f - 2 \left(1 + 2 e^2\right) \sin  f_0 + \right. \acap
\nnb & + \left. e \left[6 \left(f - f_0\right) - \ton{1 + e \cos f} \sin 2u  + \right.\right.\acap
     & + \left.\left. \left(1 + e \cos  f_0\right)\sin 2u_0 \right]}.
\end{align}
As a consequence, the gravitomagnetic shifts per orbit are
\begin{align}
\Delta p^{\ton{\rm GM}} \lb{PGM} & = 0, \acap
\Delta e^{\ton{\rm GM}} \lb{EGM} & = 0, \acap
\Delta I^{\ton{\rm GM}} \lb{IGM} & = 0, \acap
\Delta \Om^{\ton{\rm GM}} \lb{OGM} & = \rp{4\pi G S}{c^2\sqrt{p^3\mu}}, \acap
\Delta \omega^{\ton{\rm GM}} \lb{oMGM} & = -\rp{12\pi G S\cI}{c^2\sqrt{p^3\mu}}.
\end{align}
\subsection{The Newtonian quadrupole acceleration}\lb{gei2}
The acceleration due to the first even zonal harmonic coefficient $J_2$ of the expansion in multipoles of the Newtonian component of the gravitational potential of the central body is
\eqi{\bds A}^{(J_2)} = \rp{3J_2\mu R^2}{2r^4}\qua{5\bds{\hat{r}}\ton{\kap\bds\cdot\bds{\hat{r}}}^2  -2\kap\ton{\kap\bds\cdot\bds{\hat{r}}} -\bds{\hat{r}}  }.\lb{quadru}\eqf
According to Section \ref{pertu}, the radial, transverse and normal components of \rfr{quadru} for an arbitrary orientation of $\kap$ are
\begin{align}
A_R^{(J_2)}\nonumber \lb{ARJ2} &= \rp{3 J_2 \mu R^2\ton{1+e\cos f}^4}{16 p^4}\grf{
24 \kz  \sII \left(\ky  \cO-\kx  \sO\right) \sin ^2 u + \right. \\ \nonumber \\
\nonumber &+\left. 6 \cII \left[-3 \kz ^2+\left(2 \ky ^2+\kz ^2-1\right) \cOO-\right.\right.\\ \nonumber \\
\nonumber &-\left.\left. 2 \kx  \ky  \sOO+1\right] \sin ^2 u + 24 \kz  \sI  \sin 2u (\kx  \cO+\ky  \sO)+ \right. \\ \nonumber \\
\nonumber &+\left. 12 \cI  \sin 2u \left[2 \kx  \ky  \cOO+\left(2 \ky ^2+\kz ^2-1\right) \sOO\right]-\right. \\ \nonumber \\
\nonumber &-\left. \left(1 + 3 \cos 2u\right) \left[3 \kz ^2+3\left(2 \ky ^2+\kz ^2-1\right) \cOO-\right.\right. \\ \nonumber \\
          &-\left.\left. 6 \kx  \ky  \sOO-1\right]
}, \\ \nonumber \\
A_T^{(J_2)}\nonumber \lb{ATJ2} &= -\rp{3 J_2 \mu R^2\ton{1+e\cos f}^4}{8 p^4}\grf{
8 \kz  \sI\cos 2u   \left(\kx  \cO+\ky  \sO\right)+\right. \\ \nonumber \\
\nonumber &+\left. 4 \cI  \cos 2u \left[2 \kx  \ky  \cOO+\left(2 \ky ^2+\kz ^2-1\right)
\sOO\right]+ \right. \\ \nonumber \\
\nonumber & +\left. \sin 2u \left[ \sin ^2 I \left(6 \kz ^2-2\right)+\left(2 \ky ^2+\kz ^2-1\right) \left(3+\cII\right) \cOO+\right.\right.\\ \nonumber \\
          & +\left.\left. 4 \kz  \sII \left(\ky  \cO-\kx  \sO\right)-2 \kx  \ky  \left(3+\cII\right) \sOO\right]
}, \\ \nonumber \\
A_N^{(J_2)}  \nonumber \lb{ANJ2} &= -\rp{3 J_2 \mu R^2\ton{1+e\cos f}^4}{p^4}\qua{\kz  \cI +\sI  \ton{\kx  \sO-\ky  \cO}} \cdot\\ \nonumber \\
\nonumber &\cdot \qua{\cos  u  \ton{\kx  \cO+\ky  \sO}+\sin  u  \left(\kz  \sI +\right.\right. \\ \nonumber \\
&+\left.\left. \cI  \left(\ky  \cO-\kx  \sO\right)\right)}.
\end{align}
By using \rfr{Dphi}, the instantaneous shifts due to \rfr{quadru} can be obtained:  they turn out to be too cumbersome to be displayed for an arbitrary orientation of $\kap$. For the particular case $\kx=\ky=0,\kz=1$, they are
\begin{align}
\Delta p^{\ton{J_2}}\ton{f,f_0} \nnb \lb{pj2} & = \rp{J_2 R^2\sin^2 I}{2p}\grf{
e \qua{3 \cos \ton{f  +   2\omega } + \cos \ton{3 f  +   2\omega } - \right.\right.\acap
\nnb & -\left.\left. 3 \cos \ton{ f_0  +  2\omega } - \cos \ton{3  f_0  +  2\omega }} - \right.\acap
&-\left. 6 \sin \ton{f -  f_0 } \sin \ton{f  +   f_0  +  2\omega }}, \acap
\Delta e^{\ton{J_2}}\ton{f,f_0} \nnb & \lb{ej2} = \rp{ J_2 R^2}{16p^2}\grf{
\ton{\cos f  - \cos f_0 } \ton{4 \ton{5 + 7 e^2 + 7 \cos 2f   + 7 \cos 2f_0} + \right.\right.\acap
\nnb & + \left.\left. \cos f  \ton{8 \ton{7 + 5 e^2} \cos f_0  + 6 e \ton{13 + 6 \cos 2f_0  + e \cos 3f_0}} + \right.\right.\acap
\nnb & + \left.\left. e \ton{3 e \cos 4f  + 78 \cos f_0  + 6\cos 3f \ton{3 + e \cos f_0} + \right.\right.\right.\acap
\nnb & + \left.\left.\left. 20 e \cos 2f_0  + \cos 2f   \ton{6 e \cos 2f_0  + 20 e + 36 \cos f_0 } + \right.\right.\right.\acap
\nnb & + \left.\left.\left. 18\cos 3f_0  + 3 e \cos 4 f_0}} \cos 2\omega  \sin^2 I  - \right.\acap
\nnb & -\left. \ton{6 \ton{2 + 5 e^2 } \cos \ton{\frac{f  +   f_0 }{2}} + \ton{28 + 17 e^2} \cos \ton{\frac{3f  +   3f_0 }{2}} + \right.\right.\acap
\nnb & +\left.\left. 28 \ton{\cos \ton{\frac{5f + f_0}{2}} + \cos \ton{\frac{f + 5f_0}{2}}} + \right.\right.\acap
\nnb & +\left.\left. e \ton{6 \ton{10 \cos \ton{\frac{3f + f_0}{2}} + 3 \cos \ton{\frac{7f + f_0}{2}} + \right.\right.\right.\right.\acap
\nnb & + \left.\left.\left.\left. 10 \cos \ton{\frac{f + 3f_0}{2}} + 3\ton{\cos \ton{\frac{5f + 3f_0}{2}} + \right.\right.\right.\right.\right.\acap
\nnb & + \left.\left.\left.\left.\left. \cos \ton{\frac{3f + 5f_0}{2}} + \cos \ton{\frac{f + 7f_0}{2}}}} + \right.\right.\right.\acap
\nnb & +\left.\left.\left. e \ton{3 \cos \ton{\frac{5f  +   5f_0 }{2}} + 17 \cos \ton{\frac{5f + f_0}{2}} + 3 \cos \ton{\frac{9f + f_0}{2}} + \right.\right.\right.\right.\acap
\nnb & + \left.\left.\left.\left. 3 \cos \ton{\frac{7f + 3f_0}{2}} + 17 \cos \ton{\frac{f + 5f_0}{2}} + \right.\right.\right.\right.\acap
\nnb & + \left.\left.\left.\left. 3 \ton{\cos \ton{\frac{3f + 7f_0}{2}} + \cos \ton{\frac{f + 9f_0}{2}}}}}} \cdot\right.\acap
\nnb & \cdot \left. \sin^2 I\sin 2\omega\sin \ton{\frac{f -  f_0 }{2}}  + \ton{\cos f  - \cos f_0 } \ton{2 \ton{3 + e^2} + \right.\right.\acap
\nnb & + \left.\left. e \ton{6 \cos f_0  + 2 \cos f  \ton{3 + e \cos f_0} + \right.\right.\right.\acap
     & + \left.\left.\left. e \ton{\cos 2f   + \cos 2f_0 }}} \ton{1 + 3 \cII}}, \acap
\Delta I^{\ton{J_2}}\ton{f,f_0} \nnb \lb{Ij2} & =   \rp{J_2 R^2\sII}{8p^2}\grf{
e \qua{  3 \cos \ton{f  +   2\omega } - 3 \cos \ton{ f_0  +  2\omega }   + \right.\right.\acap
\nnb & +\left.\left. \cos \ton{3 f  +   2\omega } - \cos \ton{3  f_0  +  2\omega }} - \right.\acap
     & -\left. 6 \sin \ton{f -  f_0 } \sin \ton{f  +   f_0  +  2\omega }}, \acap
\Delta\Om^{\ton{J_2}}\ton{f,f_0} \nnb \lb{Oj2} & = \rp{J_2 R^2\cI}{4p^2}\grf{
 - 6 f  +  6  f_0  + 3 \sin 2u - \right.\acap
\nnb & -\left.  3 \sin 2u_0 + e \qua{ - 6 \sin f  + 6 \sin f_0  + \right.\right.\acap
\nnb & + \left.\left. 3 \sin \ton{f  +   2\omega } - 3 \sin \ton{f_0  +   2\omega } + \right.\right.\acap
     & + \left.\left.  \sin \ton{3  f  +  2\omega } - \sin \ton{3  f_0  +  2\omega }}}, \acap
\Delta\omega^{\ton{J_2}}\ton{f,f_0} \nnb  \lb{oj2} & = \rp{J_2 R^2}{64 e p^2}\grf{
120 e \ton{f -  f_0 } \cII  + 6 \ton{4  +  11 e^2} \sin f  + \right.\acap
\nnb & + \left. 8 \ton{\ton{ - 3 \sin \ton{f  +   2\omega } + 7 \sin \ton{3 f  +   2\omega } + 3 \sin \ton{ f_0  +  2\omega } - \right.\right.\right.\acap
\nnb & - \left.\left.\left. 7 \sin \ton{3  f_0  +  2\omega }} \sin^2 I  + 9 e \ton{f -  f_0 } + \right.\right.\acap
\nnb & + \left.\left. 9 \cII  \ton{\sin f  - \sin f_0 } - 3 \sin f_0 } + e \ton{12 \sin 2f  + \right.\right.\acap
\nnb & + \left.\left. 12 \ton{6 \cos \ton{2 \ton{f  +   f_0  + \omega }}\sin \ton{2f -  2f_0 } \sin^2 I  + \right.\right.\right.\acap
\nnb & + \left.\left.\left. \ton{6 \cos \ton{f  +   f_0 } \cII  + \right.\right.\right.\right.\acap
\nnb & + \left.\left.\left.\left. 2 \ton{1 - 5 \cII } \cos \ton{f  +   f_0  +  2\omega }} \sin \ton{f -  f_0 } - \sin 2f_0 } + \right.\right.\acap
\nnb & + \left.\left. e \ton{6 \ton{\sin \ton{f - 2 \omega } - \sin \ton{ f_0  - 2 \omega }} \sin^2 I  + 2 \sin 3f  - \right.\right.\right.\acap
\nnb & -\left.\left.\left. 66 \sin f_0  - 2\sin 3f_0  + 51 \sin \ton{f  +  2  I } + 3 \sin \ton{3 f  +  2  I } - \right.\right.\right.\acap
\nnb & - \left.\left.\left. 51 \sin \ton{ f_0  + 2  I } - 3 \sin \ton{3  f_0  + 2  I } + 51 \sin \ton{f - 2  I } + \right.\right.\right.\acap
\nnb & + \left.\left.\left. 3 \sin \ton{3 f - 2  I } - 51 \sin \ton{ f_0  - 2 I } - 3 \sin \ton{3  f_0  - 2  I } - \right.\right.\right.\acap
\nnb & - \left.\left.\left. 6 \cos f_0  \sin \ton{4f_0  +  2\omega} - 3 \ton{1  +  15 \cII} \sin \ton{f  +   2\omega } + \right.\right.\right.\acap
\nnb & + \left.\left.\left. \ton{3 - 19 \cII } \sin \ton{3 f  +   2\omega } + 3 \ton{\sin \ton{5 f  +   2\omega } + \right.\right.\right.\right.\acap
\nnb & + \left.\left.\left.\left. \sin \ton{ f_0  +  2\omega }} + \cII  \ton{ - 3 \sin \ton{5 f  +   2\omega } + \right.\right.\right.\right.\acap
\nnb & + \left.\left.\left.\left. 45 \sin \ton{ f_0  +  2\omega } + 19 \sin \ton{3  f_0  +  2\omega } + \right.\right.\right.\right.\acap
     & + \left.\left.\left.\left. 3 \sin \ton{5  f_0  +  2\omega }}}}}.
\end{align}

By evaluating \rfr{pj2}-\rfr{oj2} for $f=f_0 +2\pi$, one obtains the following shifts per orbit
\begin{align}
\Delta p^{\ton{J_2}} \lb{PJ2} & = 0, \acap
\Delta e^{\ton{J_2}} & = 0, \acap
\Delta I^{\ton{J_2}} & = 0, \acap
\Delta \Om^{\ton{J_2}} \lb{OJ2} & = -\rp{3\pi J_2 R^2\cI}{p^2}, \acap
\Delta \omega^{\ton{J_2}} \lb{omJ2} & = \rp{3\pi J_2 R^2\ton{3 + 5\cII}}{4p^2}.
\end{align}
\section{The mixed effects of order $J_2 c^{-2}$ and $J_2 S c^{-2}$}\lb{misti}
Here, the strategy outlined in Section \ref{gene} is applied to the perturbing accelerations of Section \ref{pertu} to analytically calculate the mixed effects of order $J_2 c^{-2}$ and $J_2 S c^{-2}$ induced by both the gravitoelectric and the gravitomagentic post-Newtonian components of the gravitational field of the rotating primary.
\subsection{The gravitoelectric effects}
For the particular case $\kx=\ky=0,\kz=1$, a straightforward but tedious calculation yields
\begin{align}
\Delta p^{(J_2~{\rm GE})}_{\rm mix}  \nonumber \lb{Dpm} &= -\rp{6\pi J_2\mu R^2\sin^2 I}{c^2 p^2}\qua{3\sin 2u_0 +\right.\\ \nonumber \\
          & + \left. 2 e^2 \soo + 3e\sin\ton{f_0+2\omega } + e\sin\ton{3f_0+2\omega} }, \\ \nonumber \\
\Delta e^{(J_2~{\rm GE})}_{\rm mix}  \nonumber \lb{Dem} &= -\rp{3\pi J_2\mu R^2\sin^2 I}{8c^2 p^3}\qua{12\sin\ton{f_0+2\omega} +28\sin\ton{3f_0+2\omega}  -\right.\\ \nonumber \\
\nonumber & - \left.  3 e^2 \sin\ton{f_0 - 2\omega} + e\ton{20 + 19 e^2}\soo +  \right.\\ \nonumber \\
\nonumber & + \left. 60 e\sin 2u_0 + 18 e\sin\ton{4f_0+2\omega} + 33 e^2 \sin\ton{f_0+2\omega}  + \right. \\ \nonumber \\
          & + \left. 17 e^2 \sin\ton{3f_0 + 2\omega}  + 3 e^2 \sin\ton{5f_0+2\omega}  }, \\ \nonumber \\
\Delta I^{(J_2~{\rm GE})}_{\rm mix}  \nonumber \lb{DIm} &= -\rp{3\pi J_2\mu R^2\sII}{2c^2 p^3}\qua{3\sin 2u_0 +\right.\\ \nonumber \\
          & + \left. 2 e^2 \soo + 3 e \sin\ton{f_0+2\omega } + e \sin\ton{3f_0+2\omega} }, \\ \nonumber \\
\Delta \Om^{(J_2~{\rm GE})}_{\rm mix}  \nonumber \lb{DOm}&=\rp{3\pi J_2\mu R^2\cI}{c^2 p^3}\qua{3\cos 2u_0 - 5 e^2 +\right.\\ \nonumber \\
\nonumber & + \left. 16e\cos f_0 +2e^2\coo +3e\cos\ton{f_0+2\omega} +\right.\\ \nonumber \\
          & + \left. e\cos\ton{3f_0 + 2\omega}   }, \\ \nonumber \\
\Delta \omega^{(J_2~{\rm GE})}_{\rm mix} \nonumber \lb{Dom} &=\rp{3\pi J_2\mu R^2}{32 c^2 e p^3}\grf{
e  \left(2  \left(91 e^2+132 \right) \cos  2I - \right.\right.\\ \nonumber \\
\nonumber & -\left.\left. e \cos  2\omega   \left(68 e  -4 e + 48 \cos  2I\left(\cos  2f_0+3\right) \cos ^3 f_0 - \right.\right.\right.\\ \nonumber \\
\nonumber & -\left.\left.\left. 36 \cos  f_0 -22 \cos  3f_0 -6 \cos  5f_0  \right)+ \right.\right.\\ \nonumber \\
\nonumber & + \left.\left. 2 e  \sin  2\omega\sin  f_0  \left(6  \cos  2I\left(\cos  4f_0 +17\right)  + \right.\right.\right.\\ \nonumber \\
\nonumber & + \left.\left.\left. 4 \left(15 \cos  2I -7\right)\cos  2f_0 - 6 \cos  4f_0 -38\right)- \right.\right.\\ \nonumber \\
\nonumber & - \left.\left. 320 e \cos \left( f_0 +2  I \right)+24 \left(3-7\cos  2I \right) \cos  2u_0 + \right.\right. \\ \nonumber \\
\nonumber & + \left.\left. 8 \sin ^2 I  \left(9 \cos \left(4  f_0 +2 \omega \right)-10 \cos  2\omega \right) \right)+ \right.\\ \nonumber \\
\nonumber & + \left. 2  \left(e  \left(57e^2 + 44 \right)+8 \sin^2 I  \left(7 \cos \left(3  f_0 +2 \omega \right)- \right.\right.\right.\\ \nonumber \\
          & -\left.\left.\left. 3 \cos \left( f_0 +2 \omega \right)\right) \right)-320 e^2 \cos \left( f_0 -2  I \right)-384 e^2\cos  f_0
}.
\end{align}
No a-priori simplifying assumptions on either $e$ or $I$ were assumed. The formulas valid for an arbitrary orientation of $\kap$ are quite cumbersome: they are explicitly displayed in Appendix \ref{appendiceA}.

The total gravitoelectric shifts per orbit of order $J_2 c^{-2}$ can be obtained by adding the indirect, mixed effects of \rfr{Dpm}-\rfr{Dom} to the direct variations induced by the post-Newtonian acceleration induced by the oblateness of the central body \cite{1988CeMec..42...81S, Sof89, 1991ercm.book.....B, 2014PhRvD..89d4043W}
\begin{align}
{\bds A}^{(J_2~{\rm GE})} \nonumber \lb{APNJ2} &=  \rp{3J_2\mu R^2}{2c^2 r^4}\qua{5\bds{\hat{r}}\ton{\kap\bds\cdot\bds{\hat{r}}}^2  -2\kap\ton{\kap\bds\cdot\bds{\hat{r}}} -\bds{\hat{r}}  }\ton{v^2 -\rp{4\mu}{r} }-\\ \nonumber \\
\nonumber &- \rp{6J_2\mu R^2}{c^2 r^4}\qua{5\ton{\bds{\hat{r}}\bds\cdot\bds v}\ton{\kap\bds\cdot\bds{\hat{r}}}^2  - 2\ton{\kap\bds\cdot\bds v}\ton{\kap\bds\cdot\bds{\hat{r}}} -\ton{\bds{\hat{r}}\bds\cdot\bds v}  }\bds v-\\ \nonumber \\
&- \rp{2J_2\mu^2 R^2}{c^2 r^5}\qua{3\ton{\kap\bds\cdot\bds{\hat{r}}}^2 - 1}\bds{\hat{r}}.
\end{align}
By using \rfr{APNJ2} into \rfr{Dphi}, evaluated for $\kx=\ky=0,\kz=1$, one obtains the direct shifts per orbit
\begin{align}
\Delta p^{(J_2~{\rm GE})}_{\rm dir} \lb{dirp} & = \rp{3\pi J_2\mu R^2 e^2\sin^2 I\soo}{c^2 p^2}, \\ \nonumber \\
\Delta e^{(J_2~{\rm GE})}_{\rm dir} & = \rp{21\pi J_2\mu R^2 e\ton{2+e^2}\sin^2 I\soo}{8c^2 p^3}, \\ \nonumber \\
\Delta I^{(J_2~{\rm GE})}_{\rm dir} & = \rp{3\pi  J_2\mu R^2 e^2\sII\soo}{4c^2 p^3}, \\ \nonumber \\
\Delta \Om^{(J_2~{\rm GE})}_{\rm dir} & = \rp{3\pi  J_2\mu R^2\cI\ton{6 - e^2\coo}}{2c^2 p^3}, \\ \nonumber \\
\Delta \omega^{(J_2~{\rm GE})}_{\rm dir} \nonumber \lb{diro} & = \rp{3\pi J_2\mu R^2}{16c^2 p^3}\grf{-32 + 3 e^2 + 2\ton{7 + 2e^2}\coo +\right.\\ \nonumber \\
       & +\left. \cII\qua{-48 + 9e^2 +2\ton{-7 + 2 e^2}\coo   } }.
\end{align}
The general expressions, valid for an arbitrary orientation of $\kap$, are displayed in Appendix \ref{appendiceB}.

As a result, the total shifts per orbit turn out to be
\begin{align}
\Delta p^{(J_2~{\rm GE})}_{\rm tot} \nonumber \lb{Dptot} & =\rp{3\pi J_2\mu R^2\sin^2 I}{c^2 p^2}\grf{6\sin 2u_0 + \right. \\ \nonumber \\
          & + \left. e\qua{5e\soo + 6\sin\ton{f_0 + 2\omega} +2\sin\ton{3f_0 + 2\omega} } }, \\ \nonumber \\
\Delta e^{(J_2~{\rm GE})}_{\rm tot} \nonumber \lb{Detot} & =-\rp{3\pi J_2\mu R^2\sin^2 I}{8c^2 p^3}\grf{12\sin\ton{f_0+2\omega} + 28\sin\ton{3f_0 + 2\omega} +\right. \\ \nonumber \\
\nonumber & +\left. e\qua{ -3e\sin\ton{f_0 - 2\omega} + 6\ton{1 + 2e^2}\soo + 60\sin 2u_0 + \right.\right.\\ \nonumber \\
\nonumber & + \left.\left. 18\sin\ton{4f_0 + 2\omega} + 33 e\sin\ton{f_0 + 2\omega} + 17 e\sin\ton{3f_0 + 2\omega}  + \right. \right.\\ \nonumber \\
          & + \left.\left. 3 e\sin\ton{5f_0 + 2\omega}     }}, \\ \nonumber \\
\Delta I^{(J_2~{\rm GE})}_{\rm tot} \nonumber \lb{DItot} & = -\rp{3\pi J_2\mu R^2\sII}{4c^2 p^3}\grf{6\sin 2u_0 +\right. \\ \nonumber \\
          & + \left. e\qua{3e\soo + 6\sin\ton{f_0 + 2\omega} + 2\sin\ton{3f_0 + 2\omega}  }  }, \\ \nonumber \\
\Delta \Om^{(J_2~{\rm GE})}_{\rm tot} \nonumber \lb{DOtot} & = \rp{3\pi J_2\mu R^2\cI}{2c^2 p^3}\grf{6-10e^2  + 6\cos 2u_0 +\right.\\ \nonumber \\
\nonumber & + \left. e\qua{ 32\cos f_0 + 3e\coo + 6\cos\ton{f_0 + 2\omega} + \right.\right.\\ \nonumber \\
          & + \left.\left. 2\cos\ton{3f_0 + 2\omega}   }  }, \\ \nonumber \\
\Delta \omega^{(J_2~{\rm GE})}_{\rm tot} \nonumber \lb{Dotot} & = \rp{3\pi J_2\mu R^2}{32c^2 p^3}\grf{4\qua{ 6 + 30e^2 +42\cII + \right.\right. \\ \nonumber \\
\nonumber & + \left.\left. 7\coo + 18\cos 2u_0   }  + \right. \\ \nonumber \\
\nonumber & + \left. \rp{1}{e}\ton{4 e^2\sin  f_0  \left(-6 \sin ^2 I \cos  4f_0  +51 \cII + \right.\right.\right.\\ \nonumber \\
\nonumber & + \left.\left.\left. 2 \ton{15 \cII -7}\cos  2f_0 -19\right)\soo  - \right.\right. \\ \nonumber \\
\nonumber & - \left.\left. 2 e\left(24 e \ton{3+\cos  2f_0} \cII  \coo  \cos ^3 f_0 + \right.\right.\right.\\ \nonumber \\
\nonumber & + \left.\left.\left. 2 e\ton{96 + 160 \cII -9\coo} \cos  f_0 - \right.\right.\right.\\ \nonumber \\
\nonumber & - \left.\left.\left.e \ton{6 e+11 \cos  3f_0 +3 \cos  5f_0}  \coo + \right.\right.\right.\\ \nonumber \\
\nonumber & + \left.\left.\left. 2 \cII  \left(-50 e^2+\left(15 e^2+7\right) \coo +42\cos 2u_0 \right)\right)  + \right.\right.\\ \nonumber \\
\nonumber & + \left.\left. 8 \sin ^2 I\left(-10 e \coo +9 e \cos \ton{4f_0 + 2\omega}-\right.\right.\right.\\ \nonumber \\
          & -\left.\left.\left. 6\cos \ton{f_0 + 2\omega}+14 \cos\ton{3f_0 + 2\omega} \right)  }}.
\end{align}
Our results can be compared with those released in\footnote{The quadrupole mass moment $Q_2$ in \cite{2014PhRvD..89d4043W} has dimensions $[Q_2]=$M L$^2$: for a direct comparisons with our results, the replacement $Q_2 \rightarrow -J_2 M R^2$ in Eq. (2.12a)-Eq.(2.12c) of \cite{2014PhRvD..89d4043W} must be made. } \cite{2014PhRvD..89d4043W} for $p,e,I$ in the case $\kx=\ky=0,\kz=1$. It turns out that \rfr{Dptot} and \rfr{DItot} agree with Eq. (2.12a) and Eq. (2.12c) of \cite{2014PhRvD..89d4043W}, respectively, and \rfr{Detot} agrees with the corrected form of Eq. (2.12b) in \cite{Willcorr}. No results for $\Om,\omega$ were released in \cite{2014PhRvD..89d4043W}.

From \rfr{Dptot}-\rfr{DItot} it turns out that the  variations of $p,e,I$ are not secular trends  because of the occurrence of the slowly varying $\omega$ as argument of the trigonometrical functions entering them. Indeed, to first order, the argument of pericenter undergoes secular precession whose dominant contribution is due to either the primary's quadrupole (\rfr{omJ2}) or the post-Newtonian Schwarzschild-like gravitoelectric field (\rfr{omegaGE}), depending on the specific astronomical system considered. As such, the shifts of $p,e,I$ average out over one full cycle of $\omega$. When the Newtonian multipolar precessions are dominant with respect to the post-Newtonian ones, it is possible to have semi-secular trends for $p,e,I$ by adopting some critical inclination scenarios \cite{1953SoSht..88....3O, 2011Ap&SS.334..115L} yielding the so called frozen-perigee configuration in which the classical pericenter precession vanishes. See the scenario proposed in Section \ref{planets}. This is not the case for $\Om$ and $\omega$ itself which, according to
\rfr{DOtot}-\rfr{Dotot}, experience secular trends because of terms not containing explicitly $\omega$. The same hold for the gravitomagnetic mixed effects calculated in Section \ref{gmmix}.
\subsection{The gravitomagnetic effects}\lb{gmmix}
The inclusion of \rfr{ALT} and \rfr{quadru} in the disturbing acceleration entering \rfr{dphidf}  yields  novel post-Newtonian mixed effects proportional to $J_2 S c^{-2}$. The resulting shifts per orbit for $\kx = \ky = 0,\kz = 1$ are
\begin{align}
\Delta p^{(J_2~{\rm GM})}_{\rm mix} \nonumber &= \frac{3 \pi GS J_2 R^2\cos  I  \sin^2 I }{c^2 \sqrt{p^5 \mu }}\left\{12 \sin 2u_0 + \right. \\ \nonumber \\
& +\left. e \left(4 \left(3\sin \ton{f_0 + 2\omega}+\sin\ton{3 f_0+2 \omega}\right)-e \sin 2\omega\right)\right\}, \\ \nonumber \\
\Delta e^{(J_2~{\rm GM})}_{\rm mix} \nonumber &= \frac{3 \pi GS J_2 R^2\cos  I  \sin^2 I }{4 c^2 \sqrt{p^7 \mu }}\left\{12 \sin \ton{f_0 + 2\omega}+\right.\\ \nonumber \\
\nonumber & + \left. 28\sin\ton{3 f_0 + 2 \omega} + e \left[-2 \left(e^2 - 10\right) \sin 2\omega + 60 \sin 2u_0 + \right.\right. \\ \nonumber \\
\nonumber & + \left.\left. 33 e \sin \ton{f_0 + 2\omega} + 17 e \sin\ton{3 f_0+2 \omega} + 18\sin \ton{4f_0 + 2\omega}+\right.\right.\\ \nonumber \\
          & + \left.\left. 3 e \sin\ton{5 f_0+2 \omega} - 3 e \sin\ton{f_0-2 \omega}\right]\right\}, \\ \nonumber \\
\Delta I^{(J_2~{\rm GM})}_{\rm mix} \nonumber &= \frac{3 \pi GS J_2 R^2\sin  I }{4 c^2 \sqrt{p^7 \mu }} \left\{ 2 \left(2 \cos f_0  \ton{9 + 11 \cos 2I} + e
\left(13\cos 2I + \right.\right.\right. \\ \nonumber \\
\nonumber &+\left.\left.\left. \cos 2f_0  \ton{7 + 9 \cos 2I} + 11\right)\right) \cos 2\omega \sin f_0 + \right.\\ \nonumber \\
\nonumber &+\left. \left(2 \cos 2f_0  \ton{9 + 11 \cos 2I} + e \cos f_0  \ton{15 + 17\cos 2I} + \right.\right. \\ \nonumber \\
& + \left.\left. e \left(\cos 3f_0  \ton{7 + 9 \cos 2I} - 2 e \cos ^2 I \right)\right) \sin 2\omega\right\}, \\ \nonumber \\
\Delta \Om^{(J_2~{\rm GM})}_{\rm mix} \nonumber &= \frac{3\pi GS  J_2   R^2 }{8 c^2 \sqrt{p^7 \mu }} \left\{2 \left(2 + e^2 - 2e \cos f_0 \right)
\ton{7 + 9\cos 2I} + \right. \\ \nonumber \\
\nonumber & + \left.  \left[e^2 - 8 e\ton{2 \cos f_0 +\cos 3f_0} - \right.\right.\acap
\nnb & -\left.\left. 20 \cos 2f_0 \right] \ton{1 + 3 \cos 2I} \cos 2\omega+\right. \\ \nonumber \\
          & + \left. 8 \left[5 \cos f_0 +e \ton{3 + 2\cos 2f_0}\right] \ton{1 + 3 \cos 2I} \sin f_0  \sin 2\omega\right\}, \\ \nonumber \\
\Delta \omega^{(J_2~{\rm GM})}_{\rm mix} \nonumber &= -\frac{3 \pi GS J_2 R^2\cos  I}{16 c^2 e \sqrt{p^7 \mu }}  \left\{-56 e^2\cos f_0  -100 e^2\cos\ton{f_0-2 I}+\right.\\ \nonumber \\
\nonumber & + \left. e\left(8 \left(9\cos \ton{4f_0 + 2\omega}-10 \cos 2\omega\right) \sin^2 I + \right.\right.\\ \nonumber \\
\nonumber & + \left.\left. 4 \left(64 + 21 e^2\right) \cos 2I - e \cos\ton{f_0 + 2 I}+8\ton{17 - 37 \cos 2I}\cos 2u_0 - \right.\right. \\ \nonumber \\
\nonumber & - \left.\left. 2 e \cos 2\omega \left(-6 \sin^2 I\cos 5f_0  + 3 e - \right.\right.\right.\\ \nonumber \\
\nonumber & - \left.\left.\left. 46 \cos f_0 - 23 \cos 3f_0 - 7 e \cos 2I + \right.\right.\right.\\ \nonumber \\
\nonumber & + \left.\left.\left. 55 \ton{2 \cos f_0 + \cos 3 f_0} \cos 2I \right) + 2 e \left(-6 \ton{13 + \cos 4f_0} + \right.\right.\right.\\ \nonumber \\
\nonumber & + \left.\left.\left.  6 \ton{29 + \cos 4 f_0} \cos 2I + \right.\right.\right. \\ \nonumber \\
\nonumber & + \left.\left.\left. 4 \cos 2f_0  \ton{29 \cos 2I -13}\right)\sin f_0 \sin 2\omega\right) + \right.\\ \nonumber \\
\nonumber & + \left. 4 \left(4 \left(7 \cos\ton{3 f_0 + 2 \omega} - 3 \cos \ton{f_0 + 2\omega}\right) \sin^2 I + \right.\right.\\ \nonumber \\
          & + \left.\left. e \left(40 + 11 e^2\right)\right)\right\}.
\end{align}
We do not shown the full expressions for a general $\kap$: they are far too cumbersome.
\section{The Newtonian effects of order $J_2^2$}\lb{J2J2}
The general formalism of Section \ref{gene} allows us to work out also the Newtonian shifts per orbit quadratic in the oblateness of the primary. In certain scenarios of interest, they can become competitors not only of the mixed variations previously worked out but also of some of the most renown direct orbital effects.

In the special case $\kx = \ky = 0,\kz = 1$, they turn out to be
\begin{align}
\Delta p^{\left(J_2^2\right)} \nnb \lb{DpJ2J2} & = \rp{3\pi J_2^2 R^4\sin^2 I}{16 p^3}\grf{
\left[-16 e \left(3 + 5 \cII\right) \cos^3 f_0 - \right.\right.\acap
\nnb & -\left.\left.  12 \cos  2f_0  \left(3 + 5 \cII\right) + e^2 \left(13 + 15 \cII\right)\right] \soo - \right.\acap
& - \left. 8 \left[3 \cos  f_0 + e \left(2 + \cos  2f_0\right)\right] \left(3 + 5 \cII\right) \coo  \sin  f_0}, \\ \nonumber \\
\Delta e^{\left(J_2^2\right)} \nnb \lb{DeJ2J2} & = \rp{3\pi J_2^2 R^4\sin^2 I}{128 p^4}\grf{
-4 \left[3 e^2 \cos  4f_0  + 25 e^2 + 78 e\cos  f_0  + \right.\right.\acap
\nnb & + \left.\left. 18 e\cos  3f_0  + 4 \left(7 + 5 e^2\right) \cos  2f_0 + \right.\right.\acap
\nnb & + \left.\left. 20\right] \left(3 + 5 \cII\right) \coo  \sin  f_0 -\right.\acap
\nnb & - \left. 2 \left(- 26 e^3 + 9 e^2\cos  5f_0   + 54 e\cos  4f_0  + \right.\right.\acap
\nnb & + \left.\left. 60 e\cos  2f_0  \left(3 + 5 \cII\right)  + \right.\right.\acap
\nnb & + \left.\left. 80 e + 3 \left(28 + 17 e^2\right) \cos  3f_0 +5 \left(\left(28 + 17 e^2\right) \cos  3f_0 + \right.\right.\right.\acap
\nnb & + \left.\left.\left. 3 e \left(-2 e^2 + e\cos  5f_0  + 6\cos  4f_0 + 8\right)\right) \cII + \right.\right. \acap
& + \left.\left. 12 \left(1 + 3 e^2\right) \cos  f_0  \left(3 + 5 \cII\right)\right) \soo}, \\ \nonumber \\
\Delta I^{\left(J_2^2\right)} \nnb \lb{DIJ2J2} & = \rp{3\pi J_2^2 R^4\sII}{64 p^4}\grf{
\left[-16 e \left(3 + 5 \cII\right) \cos^3 f_0 - \right.\right.\acap
\nnb & - \left.\left. 12 \left(3 + 5 \cII\right) \cos  2f_0  + e^2 \left(13 + 15 \cII\right)\right] \soo - \right.\acap
     & - \left. 8 \left[3 \cos  f_0 + e \left(2 + \cos  2f_0 \right)\right] \left(3 + 5 \cII\right) \coo  \sin  f_0
}, \\ \nonumber \\
\Delta \Om^{\left(J_2^2\right)} \nnb \lb{DOJ2J2} & = -\rp{3\pi J_2^2 R^4 \cI}{32 p^4}\grf{
-2 e^2\coo  + 13 e^2 + \right.\acap
\nnb & + \left. 8 e\left[3 \cos \left( f_0 + 2\omega  \right) + \cos \left(3  f_0 + 2\omega  \right)\right]  + 24 \cos 2u_0 - \right.\acap
\nnb & - \left. 5 \cII  \left(-6 e^2\coo   + e^2 + 8 e\left(3 \cos \left( f_0 + 2\omega  \right) + \right.\right.\right.\acap
& + \left.\left.\left. \cos \left(3  f_0 + 2\omega  \right)\right)  + 24 \cos 2u_0 - 8\right) + 32
}, \\ \nonumber \\
\Delta \omega^{\left(J_2^2\right)} \nnb\lb{DoJ2J2} & = \rp{3\pi J_2^2 R^4}{512 e p^4 }\grf{
13 e \left(82 + 13 e^2\right) - \right.\acap
\nnb & - \left. 24 \cos \left( f_0 + 2\omega  \right) + 56 \cos \left(3  f_0 + 2\omega  \right) + \right.\acap
\nnb & + \left. 5 \cos  4I  \left(-9 e^3 + 6 e^2\cos \left( f_0 - 2\omega  \right)  + 6 e\left(8 + 9 e^2\right) \coo  - \right.\right. \acap
\nnb & - \left.\left. 2 e\left(132 \cos 2u_0  + 117 e \cos \left( f_0 + 2\omega  \right) + \right.\right.\right.\acap
\nnb & + \left.\left.\left. 43 e \cos \left(3  f_0 + 2\omega  \right) + 18 \cos \left(4  f_0 + 2\omega  \right)\right)  + \right.\right.\acap
\nnb & + \left.\left. 86 e + 24\cos \left( f_0 + 2\omega  \right) - 56 \cos \left(3  f_0 + 2\omega  \right)\right) + \right.\acap
\nnb & + \left. 4  \cII  \left(81 e^3 + 2 e\left(15 e \sin^2 I \cos \left(5  f_0 + 2\omega  \right) + \right.\right.\right.\acap
\nnb & + \left.\left.\left. \left(23 e^2 - 20\right) \coo - 12\cos 2u_0 - 27 e \cos \left( f_0 + 2\omega  \right) - \right.\right.\right.\acap
\nnb & - \left.\left.\left. 5 e \cos \left(3  f_0 + 2\omega  \right) + 18 \cos \left(4  f_0 + 2\omega  \right) - 3 e \cos \left( f_0 - 2\omega  \right)\right)  + \right.\right.\acap
\nnb & + \left.\left. 394 e - 24 \cos \left( f_0 + 2\omega  \right) + 56 \cos \left(3  f_0 + 2\omega  \right)\right) - \right.\acap
\nnb & - \left. 2 e \left(40 \coo + 60 \cos 2u_0  - 18 \cos \left(4  f_0 + 2\omega  \right) + \right.\right.\acap
\nnb & + \left.\left. 3 e \left(- 12 \sin^2 I \cos \left(5  f_0 + 2\omega  \right)  + e \coo + 25 \cos \left( f_0 + 2\omega  \right) + \right.\right.\right.\acap
& è \left.\left.\left. 7 \cos \left(3  f_0 + 2\omega  \right) + \cos \left( f_0 - 2\omega  \right)\right)\right)
}.
\end{align}
The formulas for an arbitrary spatial orientation of $\kap$ are too cumbersome to be explicitly displayed.

Also in this case, $p,e,I$ experience \textcolor{black}{long-period} harmonic variations because of the trigonometric functions in \rfr{DpJ2J2}-\rfr{DIJ2J2} having $\omega$ as argument. Instead, $\Om$ and $\omega$ undergo secular variations since in \rfr{DOJ2J2}-\rfr{DoJ2J2} there are terms not containing explicitly $\omega$.
\section{Phenomenological aspects of the mixed orbital effects}\lb{expe}
 In this Section, we numerically evaluate the relative strengths of the the direct and indirect shifts calculated in Section \ref{pertu}-Section \ref{J2J2} in some astronomical and astrophysical scenarios of interest.
\subsection{Compact objects}\lb{hole}
Let us start from a test particle orbiting a central compact object like, say, a BH \cite{2008ApJ...689.1044G, 2009ApJ...692.1075G, 2012Sci...338...84M} or a neutron star \cite{1992Natur.355..145W, 2003ApJ...591L.147K, 2011Sci...333.1717B, 2012NewAR..56....2W}.
The BH's angular momentum is \cite{1986bhwd.book.....S}
\eqi  S = \chi_g \rp{M^2 G}{c},\lb{SBH}\eqf
with \cite{1986bhwd.book.....S} \eqi\chi_g\leq 1.\lb{limite}\eqf
Recent measurements of the spin parameter of several BHs with a variety of techniques \cite{2003Natur.425..934G, 2009ApJ...697...45B, 2010MNRAS.403L..74K, 2011ApJ...735..110B, 2013arXiv1306.2033D, 2013Natur.494..449R, 2014Natur.507..207R} not implying the use of particle's orbital dynamics  confirm the bound of \rfr{limite}.
The existence of a maximum value for the angular momentum of a rotating BH is due to the fact that the Kerr
metric is endowed with horizons \cite{1970Natur.226...64B, 1972ApJ...178..347B}.
A value of the spin parameter larger than unity would imply the existence of a naked singularity \cite{1986bhwd.book.....S}; closed timelike curves
could be considered, implying a causality violation \cite{1983mtbh.book.....C}. The formation of naked singularities in gravitational collapse is prohibited by the cosmic censorship conjecture \cite{2002GReGr..34.1141P}, although it has not yet been demonstrated.
For other rotating astrophysical objects there is no such a limit as of \rfr{limite}.
In particular, for main--sequence stars, $\chi_g$  can be much larger than unity being strongly dependent  on the stellar mass
\cite{Kraft69, Kraft70, Dicke70, 1982ApJ...261..259G}. As far as compact stars are concerned,
in \cite{2011ApJ...728...12L} it was shown that for neutron stars with $M\gtrsim 1~\textrm{M}_{\odot}$
it should be $\chi_g \lesssim 0.7$, independently of the Equation Of State (EOS) governing the stellar matter.  Even lower values are usually admitted \cite{2013ApJ...777...68B}.
The angular momentum of hypothetical quark stars strongly depends on the EOS and the stellar mass itself in such a way that they may have $\chi_g > 1$ \cite{2011ApJ...728...12L}.

As far as a BH quadrupole mass moment is concerned, as a consequence of the \virg{no-hair} or uniqueness theorems \cite{Chru94, 1998LNP...514..157H}, all the multipole moments of the external spacetime are
functions of $M$ and $S$ \cite{1970JMP....11.2580G, 1974JMP....15...46H}. In particular, the quadrupole
moment of the BH is
\eqi Q_2 = -\rp{S^2}{c^2 M}. \lb{QBH} \eqf
In the case of spinning neutron stars, they acquire a nonzero quadrupole moment \cite{1999ApJ...512..282L, 2004MNRAS.350.1416B, 2012PhRvL.108w1104P, 2012ApJ...753..175B}
\eqi Q_2 = -q\rp{M^3 G^2}{c^4},\eqf
 where $q$ ranges from $1$ to $11$ for a variety of EOSs.

By suitably expressing the semilatus rectum  as \eqi p={\rm n}{\mathcal{R}}_s\ton{1+e},~\nbh \gg 3,\eqf  where the minimum distance of an orbiting star from the BH was written as a multiple of the BH's Schwarzschild radius, from \rfr{SBH} and \rfr{QBH} it is possible to obtain
\begin{align}
\left|\rp{\Delta p^{(J_2~{\rm GE})}}{\Delta p^{(J_2~{\rm GM})}}\right| \lb{p1} & = \rp{\sec I}{\chi_g}\sqrt{\rp{\nbh}{2}} + \mathcal{O}\ton{e},\acap
\left|\rp{\Delta p^{(J_2~{\rm GE})}}{\Delta p^{(J_2^2)}}\right| \lb{p2} & = \rp{16\nbh}{\chi^2_g\ton{3 + 5\cII}} + \mathcal{O}\ton{e},\acap
\left|\rp{\Delta p^{(J_2~{\rm GM})}}{\Delta p^{(J_2^2)}}\right| \lb{p3} & = \rp{16\cI\sqrt{2\nbh}}{\chi_g\ton{3 + 5\cII}} + \mathcal{O}\ton{e},\acap
\left|\rp{\Delta e^{(J_2~{\rm GE})}}{\Delta e^{(J_2~{\rm GM})}}\right| \lb{e1} & = \rp{\sec I}{\chi_g}\sqrt{\rp{\nbh}{2}} + \mathcal{O}\ton{e},\acap
\left|\rp{\Delta e^{(J_2~{\rm GE})}}{\Delta e^{(J_2^2)}}\right| \lb{e2} & = \rp{16\nbh}{\chi^2_g\ton{3 + 5\cII}} + \mathcal{O}\ton{e},\acap
\left|\rp{\Delta e^{(J_2~{\rm GM})}}{\Delta e^{(J_2^2)}}\right| \lb{e3} & = \rp{16\cI\sqrt{2\nbh}}{\chi_g\ton{3 + 5\cII}} + \mathcal{O}\ton{e},\acap
\left|\rp{\Delta I^{(J_2~{\rm GE})}}{\Delta I^{(J_2~{\rm GM})}}\right| \lb{I1} & = \rp{6\cI\sqrt{2\nbh}}{\chi_g\ton{9 + 11\cII}} + \mathcal{O}\ton{e},\acap
\left|\rp{\Delta I^{(J_2~{\rm GE})}}{\Delta I^{(J_2^2)}}\right| \lb{I2} & = \rp{16\nbh}{\chi_g^2\ton{3 + 5\cII}} + \mathcal{O}\ton{e},\acap
\left|\rp{\Delta I^{(J_2~{\rm GM})}}{\Delta I^{(J_2^2)}}\right| \lb{I3} & = \rp{4\sec I\ton{9 + 11\cII}\sqrt{2\nbh}}{3\chi_g\ton{3 + 5\cII}} + \mathcal{O}\ton{e}.
\end{align}
It must be recalled that $p,e,I$ do not\footnote{Actually, this is not true for an arbitrary orientation of the primary's spin axis \cite{2011PhRvD..84l4001I}.} experience first-order, direct shifts per orbit, apart from those due to \rfr{APNJ2} which were included in the overall gravitoelectric $J_2 c^{-2}$ effects. From \rfr{p1}-\rfr{I3} it can be noticed that the following hierarchy exists: $J_2^2 < (J_2~{\rm GM}) < (J_2~{\rm GE})$. For close orbits, the discrepancy among the gravitomagnetic and the gravitoelectric inclination shifts tend to reduce, as shown by \rfr{I1}.

In the case of the node $\Om$, also the direct Newtonian (\rfr{OJ2}) and post-Newtonian gravitomagnetic (\rfr{OGM}) shifts are to be taken into account. Thus, one has
\begin{align}
\left|\rp{\Delta\Om^{({\rm GM})}}{\Delta\Om^{(J_2)}}\right| & = \rp{4\sec I\sqrt{2\nbh}}{3\chi_g}+ \mathcal{O}\ton{e},\acap
\left|\rp{\Delta\Om^{({\rm GM})}}{\Delta\Om^{(J_2~{\rm GE})}}\right| & = \rp{4\sec I\sec^2\ton{f_0 + \omega}\sqrt{2\nbh^3}}{9\chi_g}+ \mathcal{O}\ton{e},\acap
\left|\rp{\Delta\Om^{({\rm GM})}}{\Delta\Om^{(J_2~{\rm GM})}}\right| \nnb & = \rp{32\nbh^2}{3\chi_g^2\qua{7 + 9\cII -5\ton{1 + 3\cII}\cos 2u_0 }}\acap
& + \mathcal{O}\ton{e},\acap
\left|\rp{\Delta\Om^{({\rm GM})}}{\Delta\Om^{(J_2^2)}}\right| \nnb & = \rp{64\sec I\sqrt{2\nbh^5}}{3\chi_g^3\qua{-4 - 5\cII +3\ton{-1 + 5\cII}\cos 2u_0 }}\acap
& + \mathcal{O}\ton{e},\acap
\left|\rp{\Delta\Om^{(J_2)}}{\Delta\Om^{(J_2~{\rm GE})}}\right| & = \rp{\sec^2\ton{f_0+\omega}\nbh}{3}+ \mathcal{O}\ton{e},\acap
\left|\rp{\Delta\Om^{(J_2)}}{\Delta\Om^{(J_2~{\rm GM})}}\right| & = \rp{4\cI\sqrt{2\nbh^3 }}{\chi_g\qua{7 + 9\cII -5\ton{1 + 3\cII}\cos 2u_0 }}+ \mathcal{O}\ton{e},\acap
\left|\rp{\Delta\Om^{(J_2)}}{\Delta\Om^{(J_2^2)}}\right| \nnb & = \rp{16\nbh^2}{\chi_g^2 \qua{-4 - 5\cII +3\ton{-1 + 5\cII}\cos 2u_0 }}\acap
& + \mathcal{O}\ton{e},\acap
\left|\rp{\Delta\Om^{(J_2~{\rm GE})}}{\Delta\Om^{(J_2~{\rm GM})}}\right| \nnb & = \rp{12\cI\cos^2\ton{f_0+\omega}\sqrt{2\nbh}}{\chi_g\qua{7 + 9\cII -5\ton{1 + 3\cII}\cos 2u_0 }}\acap
& + \mathcal{O}\ton{e},\acap
\left|\rp{\Delta\Om^{(J_2~{\rm GE})}}{\Delta\Om^{(J_2^2)}}\right| \nnb & = \rp{48\cos^2\ton{f_0+\omega}\nbh}{\chi_g^2\qua{-4 - 5\cII +3\ton{-1 + 5\cII}\cos 2u_0 }} \acap
& + \mathcal{O}\ton{e},\acap
\left|\rp{\Delta\Om^{(J_2~{\rm GM})}}{\Delta\Om^{(J_2^2)}}\right| \nnb & = \rp{2\sec I\qua{7 + 9\cII -5\ton{1 + 3\cII}\cos 2u_0 }\sqrt{2\nbh}}{\chi_g\qua{-4 - 5\cII +3\ton{-1 + 5\cII}\cos 2u_0 }} \acap
& + \mathcal{O}\ton{e}.
\end{align}
In the case of the pericenter, in addition to the same direct effects as for the node, there is also the direct, gravitoelectric shift of \rfr{omegaGE} to be taken into account. As a result, one has
\begin{align}
\left|\rp{\Delta\omega^{({\rm GE} )}}{\Delta\omega^{(J_2)}}\right| & = \rp{8\nbh}{\chi_g^2\ton{-4 + 5\sin^2 I}} + \mathcal{O}\ton{e}, \acap
\left|\rp{\Delta\omega^{({\rm GE} )}}{\Delta\omega^{({\rm GM})}}\right| & = \rp{\sec I}{\chi_g}\sqrt{\rp{\nbh}{2}} + \mathcal{O}\ton{e}, \acap
\left|\rp{\Delta\omega^{({\rm GE} )}}{\Delta\omega^{(J_2~{\rm GE })}}\right| & = \rp{16 e\nbh^2\csc^2 I }{\chi^2_g\qua{7\cos\ton{3f_0 + 2\omega} -3\cos\ton{f_0+2\omega} }} + \mathcal{O}\ton{e^2}, \acap
\left|\rp{\Delta\omega^{({\rm GE} )}}{\Delta\omega^{(J_2~{\rm GM })}}\right| & = \rp{8 e\csc^2 I\sec I\sqrt{2\nbh^5} }{\chi^3_g\qua{7\cos\ton{3f_0 + 2\omega} -3\cos\ton{f_0+2\omega} }} + \mathcal{O}\ton{e^2}, \acap
\left|\rp{\Delta\omega^{({\rm GE} )}}{\Delta\omega^{(J_2^2)}}\right| \nonumber & = \rp{256 e \nbh^3\csc^2 I }{\chi^4_g \ton{3 + 5\cII}\qua{7\cos\ton{3f_0 + 2\omega} -3\cos\ton{f_0+2\omega} }} + \acap
& + \mathcal{O}\ton{e^2}, \acap
\left|\rp{\Delta\omega^{({\rm GM})}}{\Delta\omega^{(J_2)}}\right| & = \rp{16\cI\sqrt{2\nbh}}{\chi_g\ton{3 + 5\cII}}+ \mathcal{O}\ton{e},\acap
\left|\rp{\Delta\omega^{({\rm GM})}}{\Delta\omega^{(J_2~{\rm GE})}}\right| & = \rp{16e\cot I\csc I\sqrt{2\nbh^3}}{\chi_g\qua{-3\cos\ton{f_0 + 2\omega} + 7\cos\ton{3f_0 + 2\omega} }}+ \mathcal{O}\ton{e^2},\acap
\left|\rp{\Delta\omega^{({\rm GM})}}{\Delta\omega^{(J_2~{\rm GM})}}\right|  & = \rp{16e\nbh^2\csc^2 I}{\chi_g^2\qua{-3\cos\ton{f_0 + 2\omega} + 7\cos\ton{3f_0 + 2\omega} }} + \mathcal{O}\ton{e^2},\acap
\left|\rp{\Delta\omega^{({\rm GM})}}{\Delta\omega^{(J_2^2)}}\right|  & = \rp{256e\cot I\csc I\sqrt{2\nbh^5}}{\chi_g^3\ton{3 + 5\cII}\qua{3\cos\ton{f_0 + 2\omega} - 7\cos\ton{3f_0 + 2\omega} }} + \mathcal{O}\ton{e^2},\acap
\left|\rp{\Delta\omega^{(J_2 )}}{\Delta\omega^{(J_2~{\rm GE } )}}\right| & = \rp{2e\ton{-5 + 4\csc^2 I}\nbh}{-7\cos\ton{3f_0 + 2\omega} +3\cos\ton{f_0+2\omega}}+ \mathcal{O}\ton{e^2}, \acap
\left|\rp{\Delta\omega^{(J_2 )}}{\Delta\omega^{(J_2~{\rm GM} )}}\right| & = \rp{e\ton{3 + 5\cII}\csc^2 I\sec I\sqrt{\nbh^3}}{\chi_g\sqrt{2}\qua{-7\cos\ton{3f_0 + 2\omega} +3\cos\ton{f_0+2\omega}}}+ \mathcal{O}\ton{e^2}, \acap
\left|\rp{\Delta\omega^{(J_2 )}}{\Delta\omega^{(J_2^2)}}\right| & = \rp{16e\nbh^2\csc^2 I }{\chi_g^2 \qua{-7\cos\ton{3f_0 + 2\omega} +3\cos\ton{f_0+2\omega}}} + \mathcal{O}\ton{e^2}, \acap
\left|\rp{\Delta\omega^{(J_2~{\rm GE} )}}{\Delta\omega^{(J_2~{\rm GM } )}}\right| & = \rp{\sec I}{\chi_g}\sqrt{\rp{\nbh}{2}} + \mathcal{O}\ton{e}, \acap
\left|\rp{\Delta\omega^{(J_2~{\rm GE} )}}{\Delta\omega^{(J_2^2 )}}\right| & = \rp{16\nbh}{\chi^2_g\ton{3 + 5\cos 2I}} + \mathcal{O}\ton{e}, \acap
\left|\rp{\Delta\omega^{(J_2~{\rm GM} )}}{\Delta\omega^{(J_2^2 )}}\right| & = \rp{16\cI\sqrt{2\nbh}}{\chi_g\ton{3 + 5\cos 2I}} + \mathcal{O}\ton{e}.
\end{align}

It must remarked that the previous expressions hold in coordinate system whose reference $\grf{x,y}$ plane coincides with the equatorial plane of the central body. In general, this is not true for BHs because the current uncertainties in the spatial orientation of their spin axes  \cite{2003A&A...412L..61K, 2013MNRAS.429L..30L, 2013ApJ...779..154L}. In hypothetical binary systems made of a BH orbited by a radiopulsar (PSR-BH) \cite{1991ApJ...379L..17N}, useful information on the magnitude and orientation of the BH's spin can be derived, in principle,  from the binary's orbital precession \cite{1999ApJ...514..388W} . As such, an accurate sensitivity analysis or error budget for some realistic scenarios require to use the fully general expressions, not displayed here because of their cumbersomeness. In the case of the Solar System, the equatorial plane of the Sun does not coincide with, say, the ecliptic plane; for a transition from one to another see, e.g., \cite{2011MNRAS.415.3335X, 2013osos.book.....X}.
\subsection{Planet-spacecraft scenarios}\lb{planets}
In this Section, we will consider some spacecraft-based scenarios in which the primary is a planet of our Solar system.

In Table \ref{tavolaJuno}, we look at Jupiter and the Juno mission \cite{2007AcAau..61..932M}, which seems promising for testing some aspects of post-Newtonian gravity \cite{2010NewA...15..554I, 2011AGUFM.P41B1620F, 2011Icar..216..440H, 2013CQGra..30s5011I}. We numerically maximized the various shifts per orbit viewed as functions of $f_0,\omega$.
\begin{table}[!t]
\caption{Maximum nominal values for the direct and mixed shifts per orbit of the Jupiter-Juno system as functions of $f_0,~\omega$. The relevant physical parameters for the giant planet are $\mu = 1.267\times 10^{17}$ m$^3$ s$^{-2}$, $R=71,492$ km, $S = 6.9\times 10^{38}$ kg m$^2$ s$^{-1}$, $J_2 = 0.014$, while for Juno we adopted $a=20.03~R$, $P_{\rm b}= 11$ d, $e=0.947,~I=90.05$ deg. The figures quoted hold in a planetary equatorial coordinate system.}\lb{tavolaJuno}
\smallskip
\begin{tabular}{llllll}
\hline\noalign{\smallskip}
& $p$ (m) & $e$ (mas) & $I$ (mas) & $\Om$ (mas) & $\omega$ (mas)\\
\noalign{\smallskip}\hline\noalign{\smallskip}
GE & $-$ & $-$ & $-$ & $-$ & $ 37.095 $  \\
GM & $-$ & $-$ & $-$ & $ 2.07 $  & $ 0.005 $ \\
$J_2$ & $-$ &  $-$ & $-$ & $ 5,835.93 $ & $ 3\times 10^6 $ \\
$J_2 c^{-2}$ & $ 0.74 $ & $ 1.4 $  & $0.0004 $ & $ 0.0012 $ & $ 1.37 $ \\
$J_2 Sc^{-2}$ & $ 0.000087 $ & $ 0.0002 $ &  $ 0.010 $ & $ 0.058 $ & $0.00027 $ \\
$J_2^2$ & $ 54,088.8 $ &  $ 130,360$ & $ 32.94 $ & $ 191.517 $ & $ 144,885 $ \\
\noalign{\smallskip}\hline\noalign{\smallskip}
\end{tabular}
\end{table}
It can be noticed that the nominal values of the Newtonian shifts of order $\mathcal{O}\ton{J_2^2}$ are far not negligible: they must be carefully accounted for in accurate error budgets when their mismodeling has to be evaluated.
The mixed post-Newtonian effects proportional to $J_2 S c^{-2}$ are quite small, being at the level of about 90 $\mu$m per orbit as far as the semilatus rectum is concerned; the angular orbital elements are shifted by far less than one milliarcsecond (mas).
The impact of the shifts of order $\mathcal{O}\ton{J_2 c^{-2}}$ is at the level of about $70$ cm per orbit ($p$), and of about 1 mas or less for the other elements.

Table \ref{tavolaLARES} shows the results for the recently launched terrestrial geodetic satellite LARES \cite{2011AcAau..69..127P, 2013NewA...23...63R}.
\begin{table}[!t]
\caption{Maximum nominal values for the direct and mixed shifts per orbit of the Earth-LARES system as functions of $f_0,~\omega$. The relevant physical parameters for our planet are $\mu = 3.986\times 10^{14}$ m$^3$ s$^{-2}$, $R=6,378$ km, $S = 5.86\times 10^{33}$ kg m$^2$ s$^{-1}$, $J_2 =0.00108$, while for LARES we adopted $a=7,826$ km, $P_{\rm b} = 1.91$ hr, $e=0.000825,~I=69.49$ deg. The figures quoted hold in an Earth equatorial coordinate system.}\lb{tavolaLARES}
\smallskip
\begin{tabular}{llllll}
\hline\noalign{\smallskip}
& $p$ (m) & $e$ (mas) & $I$ (mas) & $\Om$ (mas) & $\omega$ (mas)\\
\noalign{\smallskip}\hline\noalign{\smallskip}
GE & $-$ & $-$ & $-$ & $-$ & $2.2$  \\
GM & $-$ & $-$ & $-$ & $0.025$  & $0.03$ \\
$J_2$ & $-$ &  $-$ & $-$ & $489,608$ & $270,067$ \\
$J_2 c^{-2}$ & $0.0001$ &  $0.0034$ & $0.0008$ & $0.001$ & $4.2$ \\
$J_2 Sc^{-2}$ & $2\times 10^{-6}$ &  $0.00004$ & $4\times 10^{-6}$ & $0.00004$ & $0.05$ \\
$J_2^2$ & $19.4$ &  $425.056$ & $95.63$ & $1,281.47$ & $516,067$ \\
\noalign{\smallskip}\hline\noalign{\smallskip}
\end{tabular}
\end{table}
The post-Newtonian quadrupolar shifts of $p$ are at the $1-100~\mu$m level per orbit, while the angular shifts are below the mas level per orbit. Also in this case, the nominal shifts of the effects of order $\mathcal{O}\ton{J_2^2}$ are not negligible, although the Earth's oblateness is currently known with a high level of accuracy.

For the sake of simplicity, let us consider a hypothetical satellite with $I=\arcsin 2/\sqrt{5}$, corresponding to either $I=63.43~\textrm{deg}$ or $I=116.56~\textrm{deg}$; from \rfr{omJ2}, it turns out that the Newtonian secular precession of the perigee due to $J_2$ vanishes. Importantly, the same holds also for the long-term Newtonian variations of the eccentricity \cite{Capde05}
and the inclination \cite{Capde05}
driven by $J_3$. From, say, \rfr{Dptot}, the vanishing of $\Delta\omega^{(J_2)}$ implies that the signature of $p$ essentially looks like an almost\footnote{\textcolor{black}{The much smaller Schwarzschild-type gravitoelectric perigee variation $\Delta \omega^{\ton{\rm GE}}$ of \rfr{omegaGE} does not vanish because it is independent of $I$.}} secular trend over an observational time span of just a few years. According to \rfr{DpJ2J2}, the same is generally true also for the Newtonian shift quadratic in $J_2$. It turns out that it is possible to suitably select the initial conditions for $f_0,\omega_0$ in order to make the nominal Newtonian signature of \rfr{DpJ2J2} much smaller than the post-Newtonian one of \rfr{Dptot}. By using the values of, say, LARES for $a,e$, one obtains that $p$ experiences a post-Newtonian gravitoelectric \textcolor{black}{semi-secular shift} as large as
\eqi \dot p^{(J_2~{\rm GE})} = 51~\textrm{cm}~\textrm{yr}^{-1}~(f_0 = 1.24\times 10^{-10}~\textrm{deg},~\omega_0 = 205.258~\textrm{deg}),\lb{super}\eqf
while the competing Newtonian signal of order $\mathcal{O}\ton{J_2^2}$ essentially vanishes.
The result of \rfr{super}, whose magnitude could be increased by allowing for a more eccentric orbit, is quite large for the present-day possibilities; indeed, recent data analysis of just one year of LARES observations, processed with up-to-date models of non-gravitational perturbations, exhibited an ability to detect secular trends in $p\simeq a$ down to a $14~\textrm{cm}~\textrm{yr}^{-1}$ accuracy level \cite{Sos2013}. The same reasonings applied to \rfr{Detot}-\rfr{DItot} and to \rfr{DeJ2J2}-\rfr{DIJ2J2} yield \textcolor{black}{shifts} for $e$ and $I$ \textcolor{black}{of the order of}
\begin{align}
\dot e^{(J2~\textrm{GE})} & = -11.2~\textrm{mas}~\textrm{yr}^{-1}~(f_0 = 1.23\times 10^{-10}~\textrm{deg},~\omega_0 = 205.078~\textrm{deg}),\acap
\dot I^{(J2~\textrm{GE})} & = -3.3 ~\textrm{mas}~\textrm{yr}^{-1}~(f_0 = 8.60\times 10^{-8}~\textrm{deg},~\omega_0 = 205.069~\textrm{deg}),
\end{align}
 respectively.

\textcolor{black}{As far as the primary is concerned,} no substantial competing secular perturbations of gravitational origin would affect $e$ and $I$ because, as already remarked, the critical inclination allows to cancel also the long-term harmonic shifts due to $J_3$. \textcolor{black}{In principle, gravitational perturbations on $e,I,\Om,\omega$ arise due to the action of a third body X like, e.g., the Moon and the Sun \cite{2012CeMDA.112..117I}. Their nominal magnitude is proportional to $P_{\rm X}^{-2}P_{\rm b} =\mu_{\rm X}a_{\rm X}^{-3} P_{\rm b}$. For, say, X = Moon and a LARES-type orbit, they are of the order of $10^3$ mas yr$^{-1}$. However, since they are fully modelled, only their uncertainty, determined by the accuracy on $\mu_{\rm X}$, does matter. In the case of the Moon and the Sun, the relative accuracies in their gravitational parameters $\mu$ are several orders of magnitude better than\footnote{\textcolor{black}{They amount to \cite{2014CeMDA.119..237P} $10^{-11}$ for the Sun  and $10^{-10}$ for the Moon, respectively.}} $10^{-3}$, so that their disturbances would be negligible. } By assuming the same physical properties of LARES, the impact of the main non-gravitational perturbations \cite{1987ahl..book.....M} able to induce secular rates on $e$ and $I$ would be negligible. Indeed, according to \cite{2010AcPPB..41.4753I}, the nominal rates due to the atmospheric drag and the Rubincam effect would be as little as about $0.5$ mas yr$^{-1}$. From Eq. (6.8) of \cite{1987ahl..book.....M}, under the same assumptions as in \cite{2010AcPPB..41.4753I},  a secular decrease of the eccentricity due to the atmospheric drag as little as $0.01$ mas yr$^{-1}$ can be inferred.

Regarding the aforementioned Earth-satellite scenarios: in principle, a potential source of systematic bias may be represented by the orbital perturbations induced by the equinoctial precession \cite{2007JGCD...30..237G}. However, it must be recalled that laser data reductions are usually performed in a coordinate system whose reference $\grf{x,y}$ plane is aligned with the mean Earth's equator at the reference epoch J2000.0.
\section{Summary and conclusions}\lb{concludi}
A first-order perturbative approach to particle dynamics in the post-Newtonian  field of a rotating oblate primary is not adequate to capture the full richness of the actual orbital motion due to the simultaneous contributions of several disturbing classical and relativistic accelerations ($J_2$, Schwarzschild, Lense-Thirring, etc.). Indeed, the very same fact that more than one  enter the equations of motion induces certain indirect, mixed orbital perturbations due to a mutual cross interaction in addition to second-order effects for each of them.

A consistent formalism able to reproduce such additional features of motion, which are not directly due to some new accelerations occurring in the equations of motion, is a second-order perturbative approach which we consistently outlined and applied to some known post-Keplerian accelerations of both Newtonian and post-Newtonian origin.

In particular, we considered the Newtonian acceleration induced by the oblateness $J_2$ of the central body, and the post-Newtonian gravitoelectromagnetic accelerations of order $\mathcal{O}\ton{c^{-2}}$ which, to first order, yield the well known Einstein and Lense-Thirring orbital precessions. We analytically calculated the indirect shifts per orbit of all the standard Keplerian orbital elements proportional to $J_2 c^{-2}$ and $J_2 S c^{-2}$. Our general approach is valid for an arbitrary orientation of the primary's spin axis $\kap$.
We also considered the Newtonian second-order effects in $J_2$. As far as the indirect gravitoelectric effects of order $\mathcal{O}\ton{J_2 c^{-2}}$ are concerned, they were added to the direct ones caused by the specific post-Newtonian acceleration proportional to $J_2 c^{-2}$ entering the equations of motion.

It turned out that the semilatus rectum $p$, the eccentricity $e$ and the inclination $I$ experience non-vanishing indirect shifts which are harmonic in the  argument of pericenter $\omega$ entering their expressions as argument of trigonometric functions. The pericenter does not generally stay constant because of the direct perturbations of order $\mathcal{O}\ton{J_2}$ and $\mathcal{O}\ton{c^{-2}}$ that make it precess slowly. Instead, the node $\Om$ and the pericenter itself undergo also indirect secular precessions because of some terms not containing explicitly $\omega$. Our formulas, which are valid for a generic orbital geometry of the test particle, represent the limit to which full two-body formulas will have to reduce in the point particle limit.

Such indirect, mixed effects may play a role in realistic error budgets of accurate tests of post-Newtonian gravity and in the long-term evolutionary history of various astrophysical systems of interest. In principle, it is possible to design a dedicated satellite-based mission aimed to detect the effects of order $\mathcal{O}\ton{J_2c^{-2}}$ by looking at $p,e,I$. Indeed, in the Earth scenario, the dominant perigee precession is due to the Newtonian multipoles of the expansion of the terrestrial gravitational potential. Thus, a suitable orbital configuration, based on the concepts of critical inclination and frozen-perigee, may be adopted to suppress the largest part of the perigee precession as well as the long-term harmonic variations of the eccentricity and the inclination. In such a way, the shifts of order $\mathcal{O}\ton{J_2 c^{-2}}$ in $p,e,I$ would look like \textcolor{black}{almost} secular trends over typical observational time spans some years long \textcolor{black}{because of the very slow Schwarzschild-like gravitoelectric perigee advance}. As an example, a hypothetical terrestrial satellite orbiting at an altitude of $h=1~450$ km in an almost circular orbit inclined to the Earth's equator by an amount equal to the critical inclination able to suppress the Newtonian secular perigee precession due to $J_2$ as well as the long-term harmonic variations of the eccentricity and the inclination due to $J_3$, would experience an almost secular rate in $p$ as large as 51 cm yr$^{-1}$. Recent data analysis of the existing geodetic satellite LARES showed an accuracy in determining secular trends in the semimajor axis $a\simeq p$ of the order of 14 cm yr$^{-1}$ over just one year. The eccentricity and the inclination would change at a rate of the order of $-11$ mas yr$^{-1}$ and $-3$ mas yr$^{-1}$, respectively. The nominal magnitude of the competing rates due to the atmospheric drag are much smaller.
%
\section*{Acknowledgements}
I would like to thank M. Efroimsky, P. Gurfil and G. Xu for constructive remarks and suggestions.
\renewcommand\appendix{\par
\setcounter{section}{0}%
\setcounter{subsection}{0}%
\setcounter{table}{0}
\setcounter{figure}{0}
\setcounter{equation}{0}
\gdef\thetable{\Alph{table}}
\gdef\thefigure{\Alph{figure}}
\gdef\theequation{\Alph{section}.\arabic{equation}}
\section*{Appendix}
\gdef\thesection{\Alph{section}}
\setcounter{section}{0}}

\appendix
\section{Mixed orbital shifts of order $J_2c^{-2}$ for a generic orientation of the spin axis of the primary}\lb{appendiceA}
Here, the general expressions for the post-Newtonian gravitoelctric mixed orbital shifts arising from \rfr{Schwacc} and \rfr{quadru} are displayed for an arbitrary orientation of the primary's spin axis. In this case, the inclination $I$ does not necessarily refer to the equatorial plane of the central body, which, in general, does not coincide with the reference $\grf{x,y}$ plane. The following formulas are valid also for a general orbital configuration of the test particle.
\begin{landscape}
\small{
\begin{align}
\Delta p_{\rm mix}^{(J_2~\textrm{GE})} \nnb  \lb{bigp} & =  - \rp{3\pi J_2 \mu R^2}{2c^2 p^2}\grf{
8 \kz  \left(3 \cos  2u_0  + \right.\right.\acap
\nnb & + \left.\left. e \left(2 e \coo  + 3 \cos \left(f_0  + 2 \omega \right) + \cos \left(3 f_0  + 2 \omega \right)\right)\right) \sI  \left(\kx  \cO  + \ky  \sO \right) + \right.\acap
\nnb & + \left. 4 \kz  \sII  \left(3 \sin  2u_0  + e \left(2 e \soo  + 3 \sin \left(f_0  + 2 \omega \right) + \right.\right.\right.\acap
\nnb & + \left.\left.\left. \sin \left(3 f_0  + 2 \omega \right)\right)\right) \left(\ky  \cO  - \kx  \sO \right) + \right.\acap
\nnb & + \left. 4 \cI  \left(3 \cos  2u_0  + e \left(2 e \coo  + 3 \cos \left(f_0  + 2 \omega \right) + \cos \left(3 f_0  + 2 \omega \right)\right)\right)\cdot\right.\acap
\nnb & \cdot \left. \left(2 \kx  \ky  \cOO   + \left(2 \ky^2  + \kz^2  - 1\right) \sOO  \right) + \right.\acap
\nnb & + \left. \cII  \left(3 \sin  2u_0  + e \left(2 e \soo  + 3 \sin \left(f_0  + 2 \omega \right) + \sin \left(3 f_0  + 2 \omega \right)\right)\right)\cdot\right.\acap
\nnb & \cdot \left. \left( - 3 \kz^2  + \left(2 \ky^2  + \kz^2  - 1\right) \cOO   - 2 \kx  \ky  \sOO   + 1\right) + \right.\acap
\nnb & + \left. \left(3 \sin  2u_0  + e \left(2 e \soo  + 3 \sin \left(f_0  + 2 \omega \right) + \sin \left(3 f_0  + 2 \omega \right)\right)\right)\cdot\right.\acap
     & \cdot \left. \left(3 \kz^2  + 3 \left(2 \ky^2  + \kz^2  - 1\right) \cOO   - 6 \kx  \ky  \sOO   - 1\right)
}, \acap
\Delta e_{\rm mix}^{(J_2~\textrm{GE})} \nnb  \lb{bige} & = \rp{3\pi J_2 \mu R^2}{64c^2 p^3}\grf{
- 16 \kz  \left(4 \left(3 \cos \left(f_0  + 2 \omega \right) + 7 \cos \left(3 f_0  + 2 \omega \right)\right) + \right.\right.\acap
\nnb & + \left.\left. e \left(20 \coo  + 60 \cos  2u_0  + 18 \cos \left(4 f_0  + 2 \omega \right) + e \left(19 e \coo  + \right.\right.\right.\right.\acap
\nnb & + \left.\left.\left.\left. 33 \cos \left(f_0  + 2 \omega \right) + 17 \cos \left(3 f_0  + 2 \omega \right) + 3 \cos \left(5 f_0  + 2 \omega \right) + 3 \cos \left(f_0  - 2 \omega \right)\right)\right)\right) \sI  \left(\kx  \cO  + \ky  \sO \right) - \right.\acap
\nnb & - \left. 8 \kz  \sII  \left(4 \left(3 \sin \left(f_0  + 2 \omega \right) + 7 \sin \left(3 f_0  + 2 \omega \right)\right) + \right.\right.\acap
\nnb & + \left.\left. e \left(20 \soo  + 60 \sin  2u_0  + 18 \sin \left(4 f_0  + 2 \omega \right) + e \left(19 e \soo  + 33 \sin \left(f_0  + 2 \omega \right) + \right.\right.\right.\right.\acap
\nnb & + \left.\left.\left.\left. 17 \sin \left(3 f_0  + 2 \omega \right) + 3 \sin \left(5 f_0  + 2 \omega \right) - 3 \sin \left(f_0  - 2 \omega \right)\right)\right)\right) \left(\ky  \cO  - \kx  \sO \right) - 8 \cI  \left(4 \left(3 \cos \left(f_0  + 2 \omega \right) + \right.\right.\right.\acap
\nnb & + \left.\left.\left. 7 \cos \left(3 f_0  + 2 \omega \right)\right) + e \left(20 \coo  + 60 \cos  2u_0  + 18 \cos \left(4 f_0  + 2 \omega \right) + \right.\right.\right.\acap
\nnb & + \left.\left.\left. e \left(19 e \coo  + 33 \cos \left(f_0  + 2 \omega \right) + 17 \cos \left(3 f_0  + 2 \omega \right) + 3 \cos \left(5 f_0  + 2 \omega \right) + 3 \cos \left(f_0  - 2 \omega \right)\right)\right)\right) \left(2 \kx  \ky  \cOO   + \right.\right.\acap
\nnb & + \left.\left. \left(2 \ky^2  + \kz^2  - 1\right) \sOO  \right) - 2 \cII  \left(4 \left(3 \sin \left(f_0  + 2 \omega \right) + 7 \sin \left(3 f_0  + 2 \omega \right)\right) + e \left(20 \soo  + \right.\right.\right.\acap
\nnb & + \left.\left.\left. 60 \sin  2u_0  + 18 \sin \left(4 f_0  + 2 \omega \right) + e \left(19 e \soo  + 33 \sin \left(f_0  + 2 \omega \right) + 17 \sin \left(3 f_0  + 2 \omega \right) + 3 \sin \left(5 f_0  + 2 \omega \right) - \right.\right.\right.\right.\acap
\nnb & -\left.\left.\left.\left. 3 \sin \left(f_0  - 2 \omega \right)\right)\right)\right) \left( - 3 \kz^2  + \left(2 \ky^2  + \kz^2  - 1\right) \cOO   - 2 \kx  \ky  \sOO   + 1\right) - \right.\acap
\nnb & - \left. 2 \left(4 \left(3 \sin \left(f_0  + 2 \omega \right) + 7 \sin \left(3 f_0  + 2 \omega \right)\right) + \right.\right.\acap
\nnb & + \left.\left. e \left(20 \soo  + 60 \sin  2u_0  + 18 \sin \left(4 f_0  + 2 \omega \right) + \right.\right.\right.\acap
\nnb & + \left.\left.\left. e \left(19 e \soo  + 33 \sin \left(f_0  + 2 \omega \right) + 17 \sin \left(3 f_0  + 2 \omega \right) + 3 \sin \left(5 f_0  + 2 \omega \right) - 3 \sin \left(f_0  - 2 \omega \right)\right)\right)\right)\cdot\right.\acap
     & \cdot \left. \left(3 \kz^2  + 3 \left(2 \ky^2  + \kz^2  - 1\right) \cOO   - 6 \kx  \ky  \sOO   - 1\right)
}, \acap
\Delta I_{\rm mix}^{(J_2~\textrm{GE})} \nnb  \lb{bigI} & =  - \rp{3\pi J_2 \mu R^2}{c^2 p^3}\grf{
\left(\kz  \cI  + \sI  \left(\kx  \sO  - \ky  \cO \right)\right)\cdot\right.\acap
\nnb & \cdot \left. \left(\kz  \sI  \left(3 \sin  2u_0  + e \left(2 e \soo  + 3 \sin \left(f_0  + 2 \omega \right) + \sin \left(3 f_0  + 2 \omega \right)\right)\right) + \right.\right.\acap
\nnb & + \left.\left. \cI  \left(\ky  \cO  - \kx  \sO \right) \left(3 \sin  2u_0  + \right.\right.\right.\acap
\nnb & + \left.\left.\left. e \left(2 e \soo  + 3 \sin \left(f_0  + 2 \omega \right) + \sin \left(3 f_0  + 2 \omega \right)\right)\right) + \left(5 e^2 + \left( - 16 \cos f_0  + \right.\right.\right.\right.\acap
     & + \left.\left.\left.\left. 2 e \coo  + 3 \cos \left(f_0  + 2 \omega \right) + \cos \left(3 f_0  + 2 \omega \right)\right) e + 3 \cos  2u_0 \right) \left(\kx  \cO  + \ky  \sO \right)\right)
},\acap
\Delta \Om_{\rm mix}^{(J_2~\textrm{GE})} \nnb  \lb{bigO} & = \rp{3\pi J_2 \mu R^2}{c^2 p^3}\grf{
\csc I  \left(\kz  \cI  + \sI  \left(\kx  \sO  - \ky  \cO \right)\right)\cdot \right.\acap
\nnb & \cdot \left. \left(\kz  \left( - 5 e^2 + \left(16 \cos f_0  + 2 e \coo  + 3 \cos \left(f_0  + 2 \omega \right) + \cos \left(3 f_0  + 2 \omega \right)\right) e + \right.\right.\right.\acap
\nnb & + \left.\left.\left. 3 \cos  2u_0 \right) \sI  - \left(3 \sin  2u_0  + \right.\right.\right.\acap
\nnb & + \left.\left.\left. e \left(2 e \soo  + 3 \sin \left(f_0  + 2 \omega \right) + \sin \left(3 f_0  + 2 \omega \right)\right)\right) \left(\kx  \cO  + \ky  \sO \right) + \right.\right.\acap
\nnb & + \left.\left. \cI  \left( - 5 e^2 + \left(16 \cos f_0  + 2 e \coo  + 3 \cos \left(f_0  + 2 \omega \right) + \right.\right.\right.\right.\acap
     & + \left.\left.\left.\left. \cos \left(3 f_0  + 2 \omega \right)\right) e + 3 \cos  2u_0 \right) \left(\ky  \cO  - \kx  \sO \right)\right)
},\acap
\Delta \omega_{\rm mix}^{(J_2~\textrm{GE})} \nnb  \lb{bigo} & = \rp{3\pi J_2 \mu R^2}{64c^2 ep^3}\grf{
- 182 \cII  e^3 - 4 \coo  e^3 + \right.\acap
\nnb & +\left. 34 \cos \left(2I + 2\omega\right) e^3 + 34 \cos \left(2I - 2\omega \right) e^3 + 182 \cII  \cOO  e^3 - \right.\acap
\nnb & - \left. 140 \coo  \cOO  e^3 - 34\cos \left(2I  + 2\omega \right) \cOO  e^3 - \right.\acap
\nnb & - \left. 34 \cos \left(2I - 2\omega \right) \cOO  e^3 - 22 \cOO  e^3 + 384 \cos f_0  e^2 + \right.\acap
\nnb & + \left.  320 \cos \left(f_0 + 2I\right) e^2 + 320 \cos \left(f_0 - 2I\right) e^2 - 144 \kz^2 \cos ^3f_0  \left(\cos 2f_0  + 3\right) \cII  \coo  e^2 -\right.\acap
\nnb & - \left. 42 \cos \left(f_0  + 2 \omega \right) e^2 - 22 \cos \left(3 f_0  + 2 \omega \right) e^2 - 6 \cos \left(5 f_0  + 2 \omega \right) e^2 + \right.\acap
\nnb & + \left. 69 \cos \left(f_0 - 2I + 2 \omega \right) e^2 + 27 \cos \left(3 f_0 - 2I + 2 \omega \right) e^2 + 3 \cos \left(5 f_0 - 2I + 2 \omega \right) e^2 + \right.\acap
\nnb & + \left. 6 \cos \left(f_0  - 2 \omega \right) e^2 - 3 \cos \left(f_0 + 2I - 2 \omega \right) e^2 + 69 \cos \left(f_0  + 2I + 2\omega\right) e^2 + \right.\acap
\nnb & + \left. 27 \cos \left(3 f_0  + 2I + 2\omega\right) e^2 + 3 \cos \left(5 f_0  + 2I + 2\omega\right) e^2 - 3 \cos \left(f_0  - 2I + 2\omega\right) e^2 + \right.\acap
\nnb & + \left. 128 \cos f_0  \cOO  e^2 - 320 \cos \left(f_0 + 2I\right) \cOO  e^2 - 320 \cos \left(f_0 - 2I\right) \cOO  e^2 - \right.\acap
\nnb & - \left. 318 \cos \left(f_0  + 2 \omega \right) \cOO  e^2 - 130 \cos \left(3 f_0  + 2 \omega \right) \cOO  e^2 - \right.\acap
\nnb &-\left. 18 \cos \left(5 f_0  + 2 \omega \right) \cOO  e^2 - 69 \cos \left(f_0 - 2I + 2 \omega \right) \cOO  e^2 - 27 \cos \left(3 f_0 - 2I + 2 \omega \right) \cOO  e^2 - \right.\acap
\nnb & - \left. 3 \cos \left(5 f_0 - 2I + 2 \omega \right) \cOO  e^2 + 18 \cos \left(f_0  - 2 \omega \right) \cOO  e^2 + \right.\acap
\nnb & + \left. 3 \cos \left(f_0 + 2I - 2 \omega \right) \cOO  e^2 - 69 \cos \left(f_0  + 2I + 2\omega\right) \cOO  e^2 - \right.\acap
\nnb & -\left. 27 \cos \left(3 f_0  + 2I + 2\omega\right) \cOO  e^2 - \right.\acap
\nnb & - \left. 3 \cos \left(5 f_0  + 2I + 2\omega\right) \cOO  e^2 + 3 \cos \left(f_0  - 2I + 2\omega\right) \cOO  e^2 - \right.\acap
\nnb & - \left. 264 e\cII  + 40 \coo  e - 72 \cos  2u_0 e + \right.\acap
\nnb & + \left. 84 \cos \left(2 \left(f_0  - I + \omega \right)\right) e - 20 \cos \left(2I + 2\omega\right) e + \right.\acap
\nnb & + \left. 84 \cos \left(2 \left(f_0  + I + \omega \right)\right) e + 18 \cos \left(2 \left(2 f_0  + I + \omega \right)\right) e - \right.\acap
\nnb & - \left. 36 \cos \left(4 f_0  + 2 \omega \right) e + 18 \cos \left(4 f_0 - 2I + 2 \omega \right) e - \right.\acap
\nnb & - \left. 20 \cos \left(2I - 2\omega \right) e + 264 \cII  \cOO  e + 120 \coo  \cOO  e - \right.\acap
\nnb & - \left. 408 \cos  2u_0 \cOO  e - 84 \cos \left(2 \left(f_0  - I + \omega \right)\right) \cOO  e + 20 \cos \left(2I + 2\omega\right) \cOO  e - \right.\acap
\nnb & - \left. 84 \cos \left(2 \left(f_0  + I + \omega \right)\right) \cOO  e - 18 \cos \left(2 \left(2 f_0  + I + \omega \right)\right) \cOO  e - \right.\acap
\nnb & - \left. 108 \cos \left(4 f_0  + 2 \omega \right) \cOO  e - 18 \cos \left(4 f_0 - 2I + 2 \omega \right) \cOO  e + 20 \cos \left(2I - 2\omega \right) \cOO  e - \right.\acap
\nnb &- \left. 264 \cOO  e + 72 \cos \left(f_0  + 2 \omega \right) \cOO  - 168 \cos \left(3 f_0  + 2 \omega \right) \cOO  + \right.\acap
\nnb & + \left. 4 \ky^2 \left(160 \cos \left(f_0 - 2I\right) e^2 + 11 \left(e^2 + 12\right) e + \left( - 60 \coo  + 204 \cos  2u_0 + \right.\right.\right.\acap
\nnb & + \left.\left.\left. 54 \cos \left(4 f_0  + 2 \omega \right) + e \left(160 \cos \left(f_0 + 2I\right) + 70 e \coo  + 159 \cos \left(f_0  + 2 \omega \right) + \right.\right.\right.\right.\acap
\nnb & + \left.\left.\left.\left. 65 \cos \left(3 f_0  + 2 \omega \right) + 9 \cos \left(5 f_0  + 2 \omega \right) - 9 \cos \left(f_0  - 2 \omega \right)\right)\right) e - \right.\right.\acap
\nnb & - \left.\left. 36 \cos \left(f_0  + 2 \omega \right) + 84 \cos \left(3 f_0  + 2 \omega \right) + \right.\right.\acap
\nnb & + \left.\left. \cII  \left( - 91 e^3 + \left(\left(34 e^2 - 20\right) \coo  + 84 \cos  2u_0 + 18 \cos \left(4 f_0  + 2 \omega \right) + \right.\right.\right.\right.\acap
\nnb & + \left.\left.\left.\left. 3 e \left(23 \cos \left(f_0  + 2 \omega \right) + 9 \cos \left(3 f_0  + 2 \omega \right) + \cos \left(5 f_0  + 2 \omega \right)\right) - \right.\right.\right.\right.\acap
\nnb & - \left.\left.\left.\left. 3 e \cos \left(f_0  - 2 \omega \right)\right) e - 132 e - 12 \cos \left(f_0  + 2 \omega \right) + \right.\right.\right.\acap
\nnb & + \left.\left.\left. 28 \cos \left(3 f_0  + 2 \omega \right)\right)\right) \cOO  + 2 \kz^2 \left(160 \cos \left(f_0 - 2I\right) e^2 + 11 \left(e^2 + 12\right) e + \left( - 60 \coo  + \right.\right.\right.\acap
\nnb & + \left.\left.\left. 204 \cos  2u_0 + 54 \cos \left(4 f_0  + 2 \omega \right) + e \left(160 \cos \left(f_0 + 2I\right) + 70 e \coo  + \right.\right.\right.\right.\acap
\nnb & + \left.\left.\left.\left. 159 \cos \left(f_0  + 2 \omega \right) + 65 \cos \left(3 f_0  + 2 \omega \right) + 9 \cos \left(5 f_0  + 2 \omega \right) - \right.\right.\right.\right.\acap
\nnb & - \left.\left.\left.\left. 9 \cos \left(f_0  - 2 \omega \right)\right) + \cII  \left( - 91 e^2 + 3 \left(23 \cos \left(f_0  + 2 \omega \right) + 9 \cos \left(3 f_0  + 2 \omega \right) + \right.\right.\right.\right.\right.\acap
\nnb & + \left.\left.\left.\left.\left. \cos \left(5 f_0  + 2 \omega \right)\right) e - 3 \cos \left(f_0  - 2 \omega \right) e + \left(34 e^2 - 20\right) \coo  + \right.\right.\right.\right.\acap
\nnb & + \left.\left.\left.\left. 84 \cos  2u_0 + 18 \cos \left(4 f_0  + 2 \omega \right) - 132\right)\right) e - 12 \left(\cII  + 3\right) \cos \left(f_0  + 2 \omega \right) + \right.\right.\acap
\nnb & + \left.\left. 28 \left(\cII  + 3\right) \cos \left(3 f_0  + 2 \omega \right)\right) \cOO  + 12 \cos \left(f_0 - 2I + 2 \omega \right) \cOO  -\right.\acap
\nnb & - \left.  28 \cos \left(3 f_0 - 2I + 2 \omega \right) \cOO  + 12 \cos \left(f_0  + 2I + 2\omega\right) \cOO  - 28 \cos \left(3 f_0  + 2I + 2\omega\right) \cOO  + \right.\acap
\nnb & + \left. 2 \left(8 \left(3 \cos \left(f_0  + 2 \omega \right) - 7 \cos \left(3 f_0  + 2 \omega \right)\right) \sin^2 I  + \right.\right.\acap
\nnb & + \left.\left. e \left(57 e^2 + 44\right) \left(3 \kz^2 - 1\right)\right) + 6 \kz^2 \left(\coo  \left( - 24 \cos f_0  \sin^2 I  + 36 e \cos 2f_0  + \left(11 e^2 + 28\right) \cos 3f_0  + \right.\right.\right.\acap
\nnb & + \left.\left.\left. e \left(2 e^2 + 3 \cos 5f_0  e + 18 \cos 4f_0  - 20\right)\right) - 2 \sin f_0  \left(19 e^2 + \right.\right.\right.\acap
\nnb & + \left.\left.\left. 6 \left(6 \cos 3f_0  + e \cos 4f_0 \right) \sin^2 I  e + 54 \cos f_0  e + \right.\right.\right.\acap
\nnb & + \left.\left.\left. 14 \left(e^2 + 2\right) \cos 2f_0  + 8\right) \soo  + \cII  \left(e \left(91 e^2 - 320 \cos f_0  e + 132\right) - \right.\right.\right.\acap
\nnb & - \left.\left.\left. 2 \left(42 e \cos 2f_0  + 14 \cos 3f_0  + e \left(17 e^2 + 9 \cos 4f_0  - 10\right)\right) \coo  + \right.\right.\right.\acap
\nnb & + \left.\left.\left. 2 \left(51 e^2 + 102 \cos f_0  e + \left(30 e^2 + 28\right) \cos 2f_0  + 8\right) \sin f_0  \soo \right)\right) - \right.\acap
\nnb & - \left. 16 \kx \ky \cI  \cOO  \left(4 \left(7 \sin \left(3 f_0  + 2 \omega \right) - 3 \sin \left(f_0  + 2 \omega \right)\right) + e \left(\left(26 e^2 - 20\right) \soo  + \right.\right.\right.\acap
\nnb & + \left.\left.\left. 72 \sin  2u_0 + 57 e \sin \left(f_0  + 2 \omega \right) + 23 e \sin \left(3 f_0  + 2 \omega \right) + \right.\right.\right.\acap
\nnb & + \left.\left.\left. 18 \sin \left(4 f_0  + 2 \omega \right) + 3 e \sin \left(5 f_0  + 2 \omega \right) + 3 e \sin \left(f_0  - 2 \omega \right)\right)\right) + \right.\acap
\nnb & + \left. 8 \kx \kz \left(8 e \cO  \csc I  \left(3 \sin  2u_0 + e \left(2 e \soo  + 3 \sin \left(f_0  + 2 \omega \right) + \sin \left(3 f_0  + 2 \omega \right)\right)\right) - \right.\right.\acap
\nnb & - \left.\left. 2 \cO  \sI  \left(4 \left(7 \sin \left(3 f_0  + 2 \omega \right) - 3 \sin \left(f_0  + 2 \omega \right)\right) + \right.\right.\right.\acap
\nnb & + \left.\left.\left. e \left(\left(26 e^2 - 20\right) \soo  + 72 \sin  2u_0 + 57 e \sin \left(f_0  + 2 \omega \right) + 23 e \sin \left(3 f_0  + 2 \omega \right) + \right.\right.\right.\right.\acap
\nnb & + \left.\left.\left.\left. 18 \sin \left(4 f_0  + 2 \omega \right) + 3 e \sin \left(5 f_0  + 2 \omega \right) + \right.\right.\right.\right.\acap
\nnb & + \left.\left.\left.\left. 3 e \sin \left(f_0  - 2 \omega \right)\right)\right) + \left(\left(51 e^3 - \left(2 \left(9 e^2 - 10\right) \coo  + 60 \cos 2u_0 + \right.\right.\right.\right.\right.\acap
\nnb & + \left.\left.\left.\left.\left.  45 e \cos \left(f_0  + 2 \omega \right) + 19 e \cos \left(3 f_0  + 2 \omega \right) - 3 e \cos \left(f_0  - 2 \omega \right)\right) e + \right.\right.\right.\right.\acap
\nnb & + \left.\left.\left.\left. 132 e + 12 \cos \left(f_0  + 2 \omega \right) - 28 \cos \left(3 f_0  + 2 \omega \right) + \cII  \left( - 91 e^3 + 69 \cos \left(f_0  + 2 \omega \right) e^2 - \right.\right.\right.\right.\right.\acap
\nnb & - \left.\left.\left.\left.\left. 3 \cos \left(f_0  - 2 \omega \right) e^2 + 2 \left(17 e^2 - 10\right) \coo  e + 84 \cos  2u_0 e - 132 e -\right.\right.\right.\right.\right.\acap
\nnb & - \left.\left.\left.\left.\left.  12 \cos \left(f_0  + 2 \omega \right) + \left(27 e^2 + 28\right) \cos \left(3 f_0  + 2 \omega \right)\right)\right) \cot I  - \right.\right.\right.\acap
\nnb & - \left.\left.\left. 3 e \left(6 \cos \left(4 f_0  + 2 \omega \right) + e \cos \left(5 f_0  + 2 \omega \right)\right) \sII \right) \sO \right) + \right.\acap
\nnb & + \left. 4 \ky \kz \left(\cO  \left(\cos 3 I  \left(91 e^3 - 80 \cos f_0  e^2 - \left(34 e^3 + 3 \left(28 \cos 2f_0  + \right.\right.\right.\right.\right.\right.\acap
\nnb & + \left.\left.\left.\left.\left.\left. 9 e \cos 3f_0  + 6 \cos 4f_0  + e \cos 5f_0 \right) e - \right.\right.\right.\right.\right.\acap
\nnb & + \left.\left.\left.\left.\left. 20 e + 28 \cos 3f_0 \right) \coo \right) \csc I  - 4 \left(51 e^2 + 3 \left(6 \cos 3f_0  + e \cos 4f_0 \right) e + \right.\right.\right.\right.\acap
\nnb & + \left.\left.\left.\left. \left(30 e^2 + 28\right) \cos 2f_0  + 8\right) \sin f_0  \sII  \soo  + \cot I  \left(16 \left(3 \sin f_0  + \sin 3f_0 \right) \soo  e^2 + \right.\right.\right.\right.\acap
\nnb & + \left.\left.\left.\left. \left(16 e \cos f_0  - 11 \left(e^2 - 24 \cII  + 24\right)\right) e + \left(36 e \cos 2f_0  + \left(11 e^2 + 28\right) \cos 3f_0  + \right.\right.\right.\right.\right.\acap
\nnb & +  \left.\left.\left.\left.\left. e \left(2 e^2 + 3 \cos 5f_0  e + 18 \cos 4f_0  - 20\right)\right) \coo \right)\right) + 2 \left(8 e \csc I  \left(3 \sin  2u_0 + \right.\right.\right.\right.\acap
\nnb & + \left.\left.\left.\left. e \left(2 e \soo  + 3 \sin \left(f_0  + 2 \omega \right) + \sin \left(3 f_0  + 2 \omega \right)\right)\right) - \right.\right.\right.\acap
\nnb & - \left.\left.\left. 2 \sI  \left(4 \left(7 \sin \left(3 f_0  + 2 \omega \right) - 3 \sin \left(f_0  + 2 \omega \right)\right) + e \left( - 20 \soo  + 18 \left(4 \sin  2u_0 + \right.\right.\right.\right.\right.\right.\acap
\nnb & +  \left.\left.\left.\left.\left.\left. \sin \left(4 f_0  + 2 \omega \right)\right) + e \left(26 e \soo  + 57 \sin \left(f_0  + 2 \omega \right) + \right.\right.\right.\right.\right.\right.\acap
\nnb & + \left.\left.\left.\left.\left.\left. 23 \sin \left(3 f_0  + 2 \omega \right) + 3 \sin \left(5 f_0  + 2 \omega \right) + 3 \sin \left(f_0  - 2 \omega \right)\right)\right)\right)\right) \sO \right) + \right.\acap
\nnb & + \left. 4 \left( - 160 \kx \ky \cos \left(f_0 - 2I\right) e^2 + 160 \kx \ky \sin f_0  \sII  e^2 - \right.\right.\acap
\nnb & - \left.\left. \kx \ky \cII  \left( - 91 e^3 + \left( - 20 \coo  + 84 \cos  2u_0 + 18 \cos \left(4 f_0  + 2 \omega \right) + \right.\right.\right.\right.\acap
\nnb & + \left.\left.\left.\left. e \left(160 \cos f_0  + 34 e \coo  + 69 \cos \left(f_0  + 2 \omega \right) + 27 \cos \left(3 f_0  + 2 \omega \right) + 3 \cos \left(5 f_0  + 2 \omega \right) - \right.\right.\right.\right.\right.\acap
\nnb & - \left.\left.\left.\left.\left. 3 \cos \left(f_0  - 2 \omega \right)\right)\right) e - 132 e - 12 \cos \left(f_0  + 2 \omega \right) + 28 \cos \left(3 f_0  + 2 \omega \right)\right) - \right.\right.\acap
\nnb & - \left.\left. \kx \ky \left(11 e \left(e^2 + 12\right) - 36 \cos \left(f_0  + 2 \omega \right) + 84 \cos \left(3 f_0  + 2 \omega \right) + \right.\right.\right.\acap
\nnb & + \left.\left.\left. e \left( - 60 \coo  + 204 \cos  2u_0 + 54 \cos \left(4 f_0  + 2 \omega \right) + e \left(70 e \coo  + \right.\right.\right.\right.\right.\acap
\nnb & + \left.\left.\left.\left.\left. 159 \cos \left(f_0  + 2 \omega \right) + 65 \cos \left(3 f_0  + 2 \omega \right) + 9 \cos \left(5 f_0  + 2 \omega \right) - \right.\right.\right.\right.\right.\acap
\nnb & - \left.\left.\left.\left.\left. 9 \cos \left(f_0  - 2 \omega \right)\right)\right)\right) - 2 \left(2 \ky^2 + \kz^2 - 1\right) \cI  \left(4 \left(7 \sin \left(3 f_0  + 2 \omega \right) - \right.\right.\right.\right.\acap
\nnb & - \left.\left.\left.\left. 3 \sin \left(f_0  + 2 \omega \right)\right) + e \left( - 20 \soo  + 72 \sin  2u_0 + \right.\right.\right.\right.\acap
\nnb & + \left.\left.\left.\left. 18 \sin \left(4 f_0  + 2 \omega \right) + e \left(26 e \soo  + 57 \sin \left(f_0  + 2 \omega \right) + \right.\right.\right.\right.\right.\acap
\nnb & + \left.\left.\left.\left.\left. 23 \sin \left(3 f_0  + 2 \omega \right) + 3 \sin \left(5 f_0  + 2 \omega \right) + 3 \sin \left(f_0  - 2 \omega \right)\right)\right)\right)\right) \sOO  + \right.\acap
\nnb & + \left. 4 \cos f_0  \left(3 \kz \coo  \left(9 \kz e^2 + 4 \ky \left(7 e^2 + \left(2 - 11 e^2\right) \cII  - 2\right) \cO  \cot I \right) - \right.\right.\acap
\nnb & - \left.\left. 8 e \left(8 e \cOO  \ky^2 + 3 \kz \cO  \left(\cot I  \left(8 e + \left(13 - 17 \cII \right) \sin f_0  \soo \right) - 20 e \sII \right) \ky -\right.\right.\right.\acap
     & - \left.\left.\left. 8 e \kx \sOO  \ky + 4 e \kz^2 \left(\cOO  + 9\right) + 8 e \kx \kz \left(\cot I  - 5 \cos 3 I  \csc I \right) \sO \right)\right)
}.
\end{align}
}
\end{landscape}
It can be noted that, in the special case $\kx = \ky = 0,\kz = 1$, \rfr{bigp}-\rfr{bigo} reduce to \rfr{Dpm}-\rfr{Dom}.
\section{Direct orbital shifts of order $J_2c^{-2}$ for a generic orientation of the spin axis of the primary}\lb{appendiceB}
Here, the general expressions for the post-Newtonian gravitoelctric direct orbital shifts arising from \rfr{APNJ2} are displayed for an arbitrary orientation of the primary's spin axis. In this case, the inclination $I$ does not necessarily refer to the equatorial plane of the central body, which, in general, does not coincide with the reference $\grf{x,y}$ axis. The following formulas are valid also for a general orbital configuration of the test particle.
\begin{align}
\Delta p_{\rm dir}^{(J_2~\textrm{GE})} \nnb \lb{dirtotp} & = \rp{3\pi J_2\mu R^2 e^2}{ 4 c^2 p^2}\grf{
8 \kz  \coo  \sI  \left(\kx  \cO  + \ky  \sO \right) + \right.\acap
\nnb & + \left. 4 \cI  \coo  \left[2 \kx  \ky  \cOO  + \left(\ky^2 - \kx^2\right) \sOO \right] + \right.\acap
\nnb & + \left. \soo  \left[\left(6 \kz^2 - 2\right) \sin^2 I + \right.\right.\acap
\nnb & + \left.\left. \left(2 \ky^2 + \kz^2 - 1\right) \left(3 + \cII\right) \cOO  + \right.\right.\acap
\nnb & + \left.\left. 4 \kz  \sII  \left(\ky  \cO  - \kx  \sO \right) - \right.\right.\acap
     & - \left.\left. 2 \kx  \ky  \left(3 + \cII\right) \sOO \right]
},\acap
\Delta e_{\rm dir}^{(J_2~\textrm{GE})} \nnb & = \rp{21\pi J_2\mu R^2 e\ton{2 + e^2}}{32 c^2 p^3}\grf{ 8\kz\sI\coo\ton{\kx\cO + \ky\sO}  +\right.\\ \nonumber \\
\nonumber &  + \left. 4\cI\coo\qua{2\kx\ky\cOO - \ton{\kx^2 - \ky^2}\sOO } - \right. \\ \nonumber \\
\nonumber & - \left. \soo\qua{ \ton{\kx^2 - \ky^2}\ton{3 + \cII}\cOO +2\ton{1-3\kz^2}\sin^2 I -\right.\right. \\ \nonumber \\
\nonumber & - \left.\left. 4\kz\sII\ton{\ky\cO-\kx\sO} + \right.\right. \\ \nonumber \\
 & + \left.\left. 2\kx\ky\ton{3 + \cII}\sOO   }  }, \\ \nonumber \\
\Delta I_{\rm dir}^{(J_2~\textrm{GE})} \nnb & = \rp{3\pi J_2\mu R^2}{2 c^2 p^3}\qua{\kz\cI +\sI\ton{\kx\sO - \ky\cO} }\\ \nonumber \\
\nonumber &\qua{e^2\kz\sI\soo +e^2\cI\soo\ton{\ky\cO-\kx\sO} + \right.\\ \nonumber \\
& + \left. \ton{6 + e^2\coo}\ton{\kx\cO + \ky\sO}  }, \\ \nonumber \\
\Delta \Om_{\rm dir}^{(J_2~\textrm{GE})} \nnb & = -\rp{3\pi J_2\mu R^2\csc I}{2 c^2 p^3}\qua{\kz\cI +\sI\ton{\kx\sO - \ky\cO} }\\ \nonumber \\
\nonumber &\grf{\ton{-6 + e^2\coo} \qua{\kz\sI+\cI\ton{\ky\cO -\kx\sO}} - \right. \\ \nonumber \\
& -\left. e^2\soo\ton{\kx\cO+\ky\sO}  }, \\ \nonumber \\
\Delta \omega_{\rm dir}^{(J_2~\textrm{GE})} \nnb \lb{dirtoto} & = \rp{3\pi J_2\mu R^2}{32 c^2 p^3}\grf{\left(8-3 e^2\right) \left(1 - 3 \kz^2\right)+ \right. \\ \nonumber \\
\nonumber & + \left. 12 \left(3 e^2-8\right) \kz \sII \left(\kx \sO-\ky\cO\right) - \right. \\ \nonumber \\
\nonumber & -\left. 112 \kz \sI \soo (\kx \cO+\ky \sO)- \right. \\ \nonumber \\
\nonumber & - \left. 14 \coo \left[1-3\kz^2+4 \kz \sII (\kx \sO-\ky \cO)\right]+ \right. \\ \nonumber \\
\nonumber & + \left. 16 \cot I [\kz \cI+\sI(\kx \sO-\ky \cO)] \right. \\ \nonumber \\
\nonumber & \left.\left[-e^2\soo \left(\kx \cO+\ky \sO\right) + \kz \left(e^2 \coo-6\right) \sI+\right.\right. \\ \nonumber \\
\nonumber & + \left.\left. \cI \left(e^2 \coo-6\right) \left(\ky \cO-\kx \sO\right)\right]-\right. \\ \nonumber \\
\nonumber & - \left. 3 \left(3 e^2+14 \coo-8\right) \left[\ton{\kx^2 - \ky^2} \cOO+2 \kx \ky \sOO\right]+\right. \\ \nonumber \\
\nonumber & + \left.\cII \left(9 e^2-14 \coo-24\right) \left[2 \kx\ky \sOO  -1 +3 \kz^2+\right.\right. \\ \nonumber \\
\nonumber & +\left.\left. \ton{\kx^2 - \ky^2} \cOO\right]+56\cI \soo \qua{\ton{\kx^2 - \ky^2} \sOO-\right.\right. \\ \nonumber \\
& - \left.\left. 2 \kx \ky \cOO}}.
\end{align}
It can be noted that, in the special case $\kx = \ky = 0,\kz = 1$, \rfr{dirtotp}-\rfr{dirtoto} reduce to \rfr{dirp}-\rfr{diro}.
\bibliography{Gclockbib}{}

\end{document}